\documentclass[prc,superscriptaddress,unsortedaddress,twocolumn,showpacs]{revtex4}
\usepackage{graphicx}
\usepackage{amsmath}
\usepackage{amssymb}
\usepackage{times}
\usepackage{bm}
\usepackage{color}
\usepackage{ulem}

\def\fun#1#2{\lower3.6pt\vbox{\baselineskip0pt\lineskip.9pt
\def\dfrac#1#2{{\displaystyle\frac{#1}{#2}}}
\ialign{$\mathsurround=0pt#1\hfil##\hfil$\crcr#2\crcr\sim\crcr}}}
\newcommand{\beq}{\begin{equation}}
\newcommand{\eeq}{\end{equation}}
\newcommand{\bea}{\begin{eqnarray}}
\newcommand{\eea}{\end{eqnarray}}

\begin{document}

\title{Examination of the adiabatic approximation for
$(d,p)$ reactions}

\author{Yoshiki Chazono}
\email[]{chazono@rcnp.osaka-u.ac.jp}
\affiliation{Research Center for Nuclear Physics (RCNP), Osaka
University, Ibaraki 567-0047, Japan}

\author{Kazuki Yoshida}
\affiliation{Research Center for Nuclear Physics (RCNP), Osaka
University, Ibaraki 567-0047, Japan}

\author{Kazuyuki Ogata}
\affiliation{Research Center for Nuclear Physics (RCNP), Osaka
University, Ibaraki 567-0047, Japan}

\date{\today}

\begin{abstract}
\noindent
\textbf{Background:} Deuteron-induced one-neutron transfer reactions
have been
used to extract single-particle properties of nuclei, and the adiabatic
(AD) approximation is often used to simply treat the deuteron
breakup states. \\
\textbf{Purpose:} The primary goal is to examine
the validity of the AD approximation for the ($d,p$) reaction
systematically.
We clarify also the role of the closed channels often
ignored in the description of breakup reactions.
\\
\textbf{Methods:} We calculate the ($d,p$) cross sections with the
continuum-discretized coupled-channels method (CDCC) for 128
reaction systems and compare the results with those obtained by the
CDCC calculation with the AD approximation.
Effect of the closed channels are investigated by ignoring
them in CDCC. \\
\textbf{Results:} The AD approximation affects in general the
($d,p$) cross section by less than 20~\%, but some exceptional
({\it nonadiabatic}) cases for which the AD approximation
breaks down are found. The closed channels turn out to give
significant effects on the cross section at deuteron energies
less than about 10~MeV. \\
\textbf{Conclusions:} The use of the AD approximation in the
description of the ($d,p$) reaction can be justified in many cases,
with the uncertainty of less than about 20~\%. The existence of
some nonadiabatic cases nevertheless should be realized.
The neglect of the closed channels without confirming the
convergence of the CDCC result is not recommended.
\end{abstract}


\maketitle

\section{Introduction}
\label{sec1}

Nucleon transfer reactions have played a substantial role
in extracting single-particle (s.p.) properties of
nuclei. Deuteron-induced transfer reactions, that is,
A($d,p$)B and A($d,n$)C processes, are particularly important
because the s.p. information on B or C not only in the ground
state (g.s.) but also in excited states can be studied.
Furthermore, these reactions in inverse kinematics can be
applied to studies of unstable nuclei; numbers of results have
been reported in, e.g., Refs.~\cite{Jon11,Jon11a,Lee11,Pai15,Mar15}.
In these studies, the adiabatic (AD) approximation~\cite{JS70,JT74}
was employed for describing the $(p+n)$-A three-body wave function
with efficiently taking into account the breakup effect of deuteron;
this framework is called adiabatic distorted-wave approximation
(ADWA).

On the theoretical side,
the reaction mechanism of the ($d,N$) reactions ($N=p$ or $n$)
has intensively been studied with three-body reaction
theories~\cite{TJ99,Del05,Mor09,Muk14,Fuk15}.
Nowadays the calculation with
the Faddeev--Alt-Grassberger-Sandhas (FAGS)
theory~\cite{Fad60,Alt67} is feasible~\cite{DC09,Del09}
that gives the exact
solution to the ($d,N$) cross section with a given three-body
Hamiltonian;
very recently, the role of the core excitation in ($d,p$) reactions
has also been studied~\cite{Del16}.
However, situation of the ($d,N$)
reactions is still complicated; the energy dependence
of the distorting potentials for $p$ and $n$, as well as their
nonlocality has been a matter of
discussion~\cite{DC09,Del09,TJ13,TJ13a,TJ14,Bai16}.
In Refs.~\cite{TJ13,TJ13a,TJ14,Bai16} a simple prescription for
implementing these ingredients was proposed within the framework of ADWA.
This prescription is very helpful to minimize the numerical
tasks for evaluating properly ($d,N$) cross sections; its
validity depends on, however, that of the AD approximation adopted.

In this study we systematically examine the AD approximation
to the three-body scattering wave function in the initial
channel of the A($d,p$)B process. We employ
the continuum-discretized coupled-channels
method (CDCC)~\cite{Kam86,Aus87,Yah12}
as a three-body reaction model,
and compare the resulting ($d,p$) cross sections
with those calculated by CDCC with the AD approximation.
For simplicity we neglect the intrinsic spin of each nucleon in
the CDCC calculation; the zero-range approximation with
the finite-range correction~\cite{Fuk15} is adopted in the
calculation of the ($d,p$) transition matrix. Furthermore, we fix the
energy used in evaluating $p$-A and $n$-A optical potentials
at half of the incident deuteron energy;
the effect of nonlocality of the potentials are not taken into
account. We thus concentrate on the effect of the AD
approximation of the $d$-A scattering wave on the
($d,p$) cross sections.
It should be noted that, in Refs.~\cite{ND11,Upa12,PM14}, a numerical
test for ADWA has been done for some reaction systems.
In this study, we consider four target nuclei, four incident
energies, four transferred angular momenta, and two possibilities
of the neutron separation energy of the residual nucleus B;
in total we consider 128 reaction systems.
In addition to that, we investigate the effect of the
closed channels (see Sec.~\ref{sec33}) on the $(d,p)$
cross sections. The closed channels are sometimes neglected
in CDCC calculations~\cite{Upa12} and can significantly
affect reaction observables at low energies in
particular~\cite{OY16}.

The construction of this paper is as follows.
In Sec.~\ref{sec2} we briefly describe the reaction model adopted.
In Sec.~\ref{sec3} we first explain the numerical inputs
and discuss the systematics of the validity of the AD
approximation.
The role of the closed channels is also clarified.
Finally we give a summary in Sec~\ref{sec4}.

\section{Theoretical framework}
\label{sec2}

%
\begin{figure}[b]
\begin{center}
\includegraphics[width=0.4\textwidth,clip]{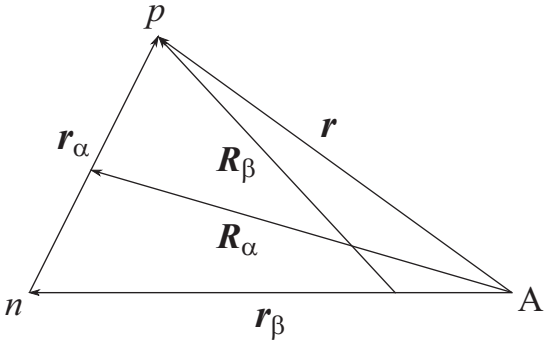}
\caption{Illustration of the three-body system.}
\label{fig1}
\end{center}
\end{figure}
We adopt the three-body system consisting of
$p$, $n$, and the target nucleus A, shown in Fig.~\ref{fig1}.
The residual nucleus B in the final channel is
assumed to be a bound state of the $n$-A system.
The post form of the transition matrix for the
A($d,p$)B process is given by
\beq
T_{\beta\alpha}=
\Big\langle
\Phi^{(-)}_\beta
\Big|
V_{pn}
\Big|
\Psi^{(+)}_\alpha
\Big\rangle,
\label{tmat}
\eeq

where $V_{pn}$ is the interaction between $p$ and $n$,
and $\Psi^{(+)}_\alpha$ is
the exact three-body scattering wave function in the initial
channel satisfying the Schr\"odinger equation
\beq
[H_\alpha-E] \Psi^{(+)}_\alpha ({\bm R}_\alpha,{\bm r}_\alpha) =0
\label{scheqi}
\eeq
with the outgoing boundary condition. The three-body Hamiltonian
$H_\alpha$ in Eq.~(\ref{scheqi}) is written as

\beq
H_\alpha=T_{{\bm R}_\alpha} + U_p (r) + U_n (r_\beta) + h_{pn},
\label{halpha}
\eeq
where $T_{{\bm R}_\alpha}$ is the kinetic energy operator regarding
the coordinate ${\bm R}_\alpha$, $U_p$ and $U_n$ are the
proton and neutron distorting potentials by A, respectively,
and $h_{pn}$ is the internal Hamiltonian of the $p$-$n$ system.
Definition of $\Phi^{(-)}_\beta$ is given below.

We adopt CDCC to obtain $\Psi^{(+)}_\alpha$:
\beq
\Psi^{(+)}_\alpha=\sum_{i=0}^{i_{\rm max}}
\phi_i ({\bm r}_\alpha) \chi_i^{(+)} ({\bm R}_\alpha),
\label{cdcc}
\eeq
where $\phi_0$ is the deuteron bound-state wave function and
$\phi_i$ for $i\neq 0$ denote discretized continuum states.
The $\phi_i$ satisfy
\beq
h_{pn} \phi_i ({\bm r}_\alpha) = \epsilon_i \phi_i ({\bm r}_\alpha)
\eeq
with $\epsilon_i$ being the eigenenergy of the $p$-$n$ system.
Equation~(\ref{cdcc}) means that the total wave function is
expanded in terms of the set of the eigenstates of $h_{pn}$,
which is assumed to form a complete set in the space relevant
to the physics observables of interest. The expansion
{\lq\lq}coefficients'' are denoted by $\chi_i^{(+)}$ which physically
represent the scattering waves between A and the $p$-$n$ system
in the $i$th state.
Although CDCC is not an exact theory for three-body scattering
processes, its theoretical foundation 
is given in Refs.~\cite{Aus89,Aus96} in connection with the
distorted-wave Faddeev formalism~\cite{BR82},
and thus can be regarded as a very good approximation to the
FAGS theory~\cite{Fad60,Alt67}.
It should be noted that the striking difference between
the results of CDCC and FAGS for low-energy deuteron breakup cross
sections found in Ref.~\cite{Upa12} was shown to be mainly because
of the lack of the CDCC model space~\cite{OY16}.
In Ref.~\cite{Upa12}, it was reported also that
$(d,p)$ cross sections obtained by CDCC somewhat deviate from
those by FAGS at incident deuteron energies higher than
about 40~MeV, which we do not discuss in this study.
For further details of CDCC,
readers are referred to Refs.~\cite{Kam86,Aus87,Yah12}.
To examine the AD approximation, we do not adopt the usual
ADWA framework but make all $\epsilon_i$
for $i \neq 0$ equal to $\epsilon_0$ in solving the CDCC equations,
to minimize the model uncertainty.
We call this calculation CDCC-AD in the following.

The three-body wave function $\Phi^{(-)}_\beta$ in the final
channel having the incoming boundary condition is a solution
of
\beq
[H_\beta-E] \Phi^{(-)}_\beta ({\bm R}_\beta,{\bm r}_\beta) =0,
\label{scheqf}
\eeq
\beq
H_\beta=T_{{\bm R}_\beta} + U_p^{\ast} (r) + h_{n{\rm A}},
\label{hbeta}
\eeq
where $T_{{\bm R}_\beta}$ is the kinetic energy operator
associated with ${\bm R}_\beta$ and $h_{n{\rm A}}$ is the
internal Hamiltonian of the $n$-A bound system.

In the present study the three-body wave function of the final
channel is approximated by
\beq
\Phi^{(-)}_\beta
\approx
\varphi_n ({\bm r}_\beta) \psi_p^{(-)} ({\bm R}_\beta),
\eeq
where $\varphi_n$ is the neutron bound-state wave-function
and $\psi_p^{(-)}$ is the distorted wave for the outgoing proton.
Because the purpose of the present study is to investigate the
validity of the AD approximation to $\Psi^{(+)}_\alpha$,
we restrict ourselves not to discuss the breakup effect
in the final channel.

The transfer reaction is described by a one-step process with
the zero-range approximation to $V_{pn} \, \phi_i$;
the finite range correction following Ref.~\cite{Fuk15} is made.
In some figures shown in Sec.~\ref{sec32}, we decompose the transition matrix
of Eq.~(\ref{tmat}) into the elastic transfer (ET) part
$T_{\beta\alpha}^{\rm ET}$ and the breakup transfer (BT) part
$T_{\beta\alpha}^{\rm BT}$ as
\bea
T_{\beta\alpha}
&=&
T_{\beta\alpha}^{\rm ET}
+
T_{\beta\alpha}^{\rm BT},
\\
T_{\beta\alpha}^{\rm ET}
&\equiv&
\Big\langle
\Phi^{(-)}_\beta
\Big|
V_{pn}
\Big|
\phi_0 ({\bm r}_\alpha) \chi_0^{(+)} ({\bm R}_\alpha)
\Big\rangle,
\label{tmatet}
\\
T_{\beta\alpha}^{\rm BT}
&\equiv&
\Big\langle
\Phi^{(-)}_\beta
\Big|
V_{pn}
\Big|
\sum_{i\neq 0}^{i_{\rm max}}
\phi_i ({\bm r}_\alpha) \chi_i^{(+)} ({\bm R}_\alpha)
\Big\rangle.
\label{tmatbt}
\eea
The cross section calculated with replacing
$T_{\beta\alpha}$ with $T_{\beta\alpha}^{\rm ET}$ ($T_{\beta\alpha}^{\rm BT}$)
is designated as the ET (BT) cross section.

\section{Results and discussion}
\label{sec3}

\subsection{Numerical inputs}
\label{sec31}

We consider four target nuclei having the atomic number $Z$
and the mass number $A$ of $(Z,A)=(10,20)$, $(20,40)$, $(40,100)$,
and $(80,200)$, which we call in the following
$^{20}$Ne, $^{40}$Ca, $^{100}$Zr, and $^{200}$Hg, respectively. These nuclei
are assumed to have a fictitious s.p. structure
so that a neutron is transferred to $\ell_f=0$, 1, 2, or 3 orbits in the
residual nucleus B, where $\ell_f$ is
the orbital angular momentum of the transferred neutron.
The principal quantum number of the neutron
starting from 0 is determined
with the assumption that the target nucleus A has a na\"{\i}ve
shell structure; in Table~\ref{tab0} we list the s.p. orbit for
the neutron transferred to nucleus A.
Furthermore, the neutron separation energy $S_n$ of B is supposed
to be 0.1~MeV or 8.0~MeV. The Bohr-Mottelson s.p. potential~\cite{BM69}
is used to calculate the neutron bound-state wave function.
\begin{table}[hbtp]
 \caption{
 Single-particle orbit for the transferred neutron.
 }
 \label{tab0}
 \centering
 \begin{tabular}{ccccccccccc}
  
  \multicolumn{6}{c}{ } \\
  \hline \hline
  & & \multicolumn{4}{c}{$ell_{f}$} \\
  Target & & 0 & 1 & 2 & 3 \\
  \hline
  $^{20}$Ne & & $1s_{1/2}$ & $1p_{3/2}$ & $0d_{5/2}$ & $0f_{7/2}$ \\
  $^{40}$Ca & & $2s_{1/2}$ & $1p_{3/2}$ & $1d_{5/2}$ & $0f_{7/2}$ \\
  $^{100}$Zr & & $2s_{1/2}$ & $2p_{3/2}$ & $1d_{5/2}$ & $1f_{7/2}$ \\
  $^{200}$Hg & & $3s_{1/2}$ & $3p_{3/2}$ & $2d_{5/2}$ & $1f_{5/2}$ \\
  \hline \hline
  
 \end{tabular}
\end{table}

The deuteron incident energy $E_d$ is taken to be 5, 10, 20, and
40~MeV. We adopt the Koning-Delaroche (KD)~\cite{KD03} nucleon optical
potential as $U_p$ and $U_n$, and the one-range Gaussian
interaction~\cite{Ohm70} is employed as $V_{pn}$.
The $p$-$n$ discretized continuum states of the $s$- and $d$-waves,
with $k_{\rm max}=2.0$~fm$^{-1}$ and $\Delta_k=0.04$~fm$^{-1}$,
are included in CDCC, where $k_{\rm max}$ is the maximum
$p$-$n$ linear momentum (in the unit of $\hbar$) and $\Delta_k$
is the size of the momentum bin.
The CDCC equations are integrated up to
$R_\alpha=20$~fm with the increment of 0.1~fm; the Coulomb breakup is
ignored in this study.

The distorted wave $\psi_p^{(-)}$ for the outgoing proton is
calculated with the KD potential. The integration of the transition
matrix is taken up to 150 and 40~fm for $S_n=0.1$ and 8~MeV,
respectively.

\subsection{Validity of the adiabatic approximation}
\label{sec32}

In table~\ref{tab1} we show the adiabatic factor $S_{\rm AD}$
determined so as to minimize
\begin{eqnarray}
\chi^2 (S_{\rm AD})
&\equiv&
\int
\left[
\left(
\dfrac{d\sigma}{d\Omega}
- S_{\rm AD} \dfrac{d\sigma_{\rm AD}}{d\Omega}
\right)
\Big /
\left(
\dfrac{d\sigma}{d\Omega}
\right)
\right]^2
\nonumber \\
&&
\times
\Theta
\left(
\dfrac{d\sigma}{d\Omega}
- \frac{1}{2}\dfrac{d\sigma^{\rm max}}{d\Omega}
\right)
\label{sad}
d\theta_{\rm cm},
\end{eqnarray}
where $d\sigma/d\Omega$ and $d\sigma_{\rm AD}/d\Omega$
are the $(d,p)$ differential cross sections calculated with CDCC
and CDCC-AD, respectively, $\Theta$ is the step function,
and $d\sigma^{\rm max}/d\Omega$ is the maximum value of $d\sigma/d\Omega$.
It should be noted that in the integration in Eq.~(\ref{sad})
we ignore the weighting factor $\sin \theta$,
where $\theta$ is the scattering angle of the outgoing
proton in the center-of-mass (c.m.) frame, as in the
standard $\chi^2$-fitting procedure for the angular distribution.

\begin{table}[hbtp]
 \caption{
Adiabatic factor $S_{\rm AD}$.
The superscripts $*1$, $*2$, and $*3$ indicate the cases in which
the AD approximation does not work. See the text for details.
}
 \label{tab1}
 \centering
 \begin{tabular}{ccccccccccc}

  \multicolumn{11}{c}{ } \\
  \multicolumn{11}{l}{$ell_f=0$} \\
  \hline \hline
  & & \multicolumn{4}{c}{Energy ($S_n=0.1$ MeV)} & & \multicolumn{4}{c}{Energy ($S_n=8$ MeV)} \\
  Target & & 5 & 10 & 20 & 40 & & 5 & 10 & 20 & 40 \\
  \hline
  $^{20}$Ne & & 0.71$^{*3}$ & 0.89 & 1.32 & 1.25 & & 0.90 & 0.93 & 0.88 & 0.74 \\
  $^{40}$Ca & & 1.08$^{*3}$ & 1.21 & 2.01$^{*2}$ & 1.44$^{*2}$  & & 0.78 & 0.77 & 0.78 & 0.87 \\
  $^{100}$Zr & & 0.96 & 1.11 & 1.83$^{*2}$ & 1.67$^{*2}$ & & 0.87$^{*1}$ & 0.87 & 0.68 & 1.10 \\
  $^{200}$Hg & & 1.00 & 0.94 & 1.21 & 1.29 & & 1.06$^{*1}$ & 0.88 & 0.66 & 1.24 \\
  \hline \hline

  \multicolumn{11}{c}{ } \\
  \multicolumn{11}{l}{$ell_f=1$} \\
  \hline \hline
  & & \multicolumn{4}{c}{Energy ($S_n=0.1$ MeV)} & & \multicolumn{4}{c}{Energy ($S_n=8$ MeV)} \\
  Target & & 5 & 10 & 20 & 40 & & 5 & 10 & 20 & 40 \\
  \hline
  $^{20}$Ne & & 0.94 & 1.00 & 0.99 & 0.92 & & 0.84 & 0.83 & 0.83 & 0.91 \\
  $^{40}$Ca & & 0.94 & 0.80 & 0.87 & 1.15 & & 0.81$^{*1}$ & 0.80 & 0.93 & 0.85 \\
  $^{100}$Zr & & 0.96 & 0.90 & 0.74 & 2.00$^{*2}$ & & 0.96 & 0.80 & 0.85 & 1.02 \\
  $^{200}$Hg & & 1.00 & 0.93 & 1.06 & 1.56$^{*2}$ & & 0.94$^{*1}$ & 0.75$^{*1}$ & 0.74 & 0.92 \\
  \hline \hline

  \multicolumn{11}{c}{ } \\
  \multicolumn{11}{l}{$ell_f=2$} \\
  \hline \hline
  & & \multicolumn{4}{c}{Energy ($S_n=0.1$ MeV)} & & \multicolumn{4}{c}{Energy ($S_n=8$ MeV)} \\
  Target & & 5 & 10 & 20 & 40 & & 5 & 10 & 20 & 40 \\
  \hline
  $^{20}$Ne & & 0.92 & 0.82 & 0.88 & 0.98 & & 0.95 & 0.94 & 0.92 & 0.90 \\
  $^{40}$Ca & & 0.93 & 0.83 & 0.92 & 0.92 & & 0.66$^{*1}$ & 0.77$^{*1}$ & 0.84 & 0.90 \\
  $^{100}$Zr & & 0.97 & 0.85 & 0.84 & 0.92 & & 0.83$^{*1}$ & 0.80$^{*1}$ & 0.77 & 0.86 \\
  $^{200}$Hg & & 1.00 & 0.93 & 0.86 & 0.97 & & 1.04$^{*1}$ & 0.76 & 0.82 & 0.88 \\
  \hline \hline

  \multicolumn{11}{c}{ } \\
  \multicolumn{11}{l}{$ell_f=3$} \\
  \hline \hline
  & & \multicolumn{4}{c}{Energy ($S_n=0.1$ MeV)} & & \multicolumn{4}{c}{Energy ($S_n=8$ MeV)} \\
  Target & & 5 & 10 & 20 & 40 & & 5 & 10 & 20 & 40 \\
  \hline
  $^{20}$Ne & & 0.89 & 0.92 & 0.92 & 0.85 & & 0.74 & 0.78 & 0.85 & 0.88 \\
  $^{40}$Ca & & 0.90 & 0.83 & 0.83 & 0.89 & & 0.87$^{*1}$ & 0.86$^{*1}$ & 0.92 & 0.98 \\
  $^{100}$Zr & & 0.98 & 0.86 & 0.82 & 0.86 & & 0.84$^{*1}$ & 0.72$^{*1}$ & 0.81 & 0.93 \\
  $^{200}$Hg & & 1.00 & 0.93 & 0.82 & 0.83 & & 0.99 & 0.75$^{*1}$ & 0.74 & 0.74 \\
  \hline \hline

 \end{tabular}
\end{table}
%
One sees from Table~\ref{tab1} that $S_{\rm AD}$ does not largely
deviate from unity in general; the AD approximation affects the
($d,p$) cross section by less than 20~\% and by about 35~\% at most.
In some exceptional cases, however, $S_{\rm AD}$ has a very large value,
meaning the clear breakdown of the AD approximation. Furthermore, there
are some cases in which $S_{\rm AD}$ is quite close to unity
but the angular distribution of the transfer
cross section is severely affected by the AD approximation.
The angular distribution of the ($d,p$) cross sections
for the 128 systems calculated with CDCC and CDCC-AD
can be found in the Addendum provided as supplemental material~\cite{SM17}.

Before discussing the {\it nonadiabatic} cases one by one,
let us first see typical cases in which the AD approximation
works well.
\begin{figure}[htpb]
\begin{center}
 \includegraphics[width=0.42\textwidth]{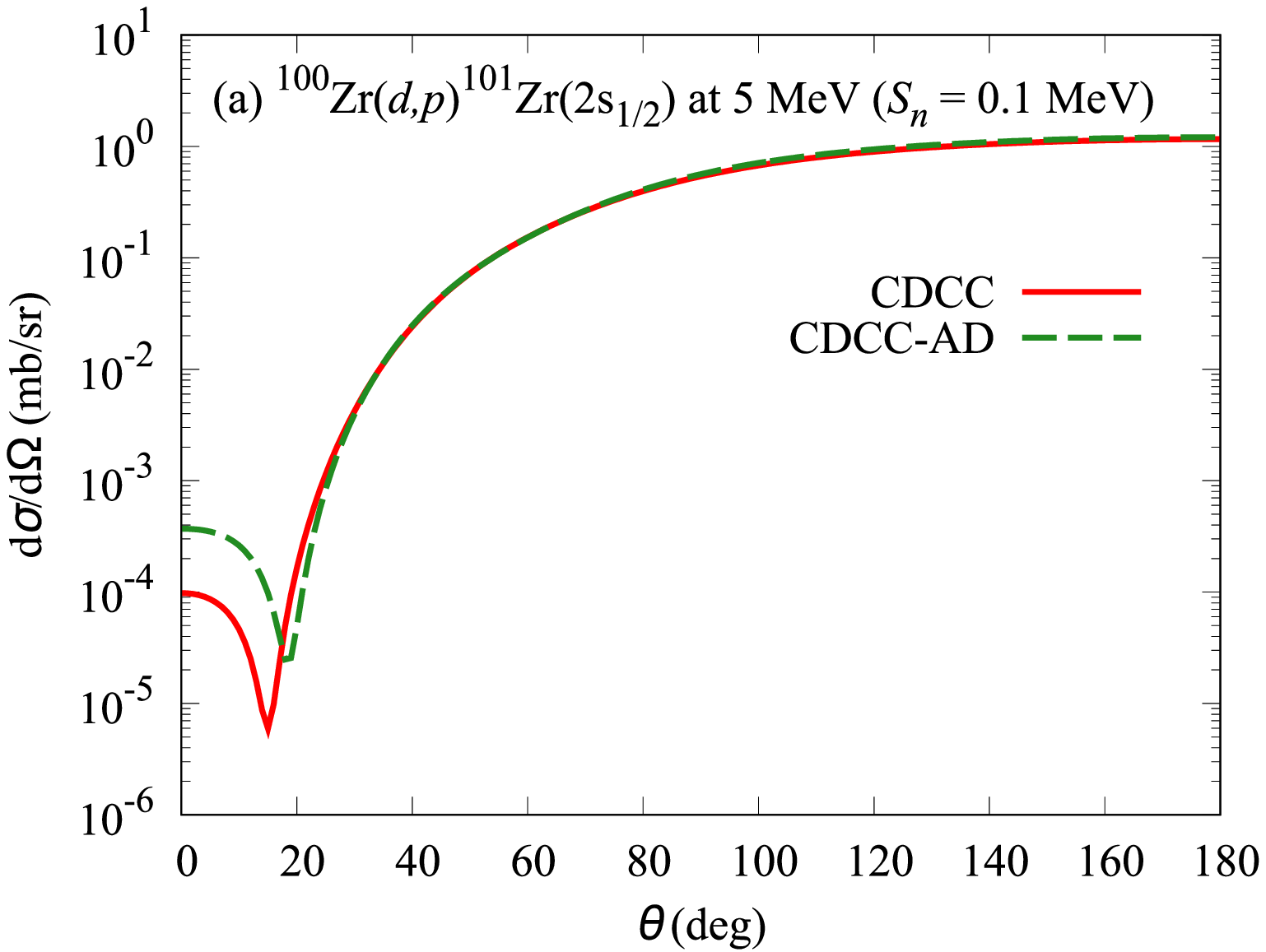}
 \includegraphics[width=0.42\textwidth]{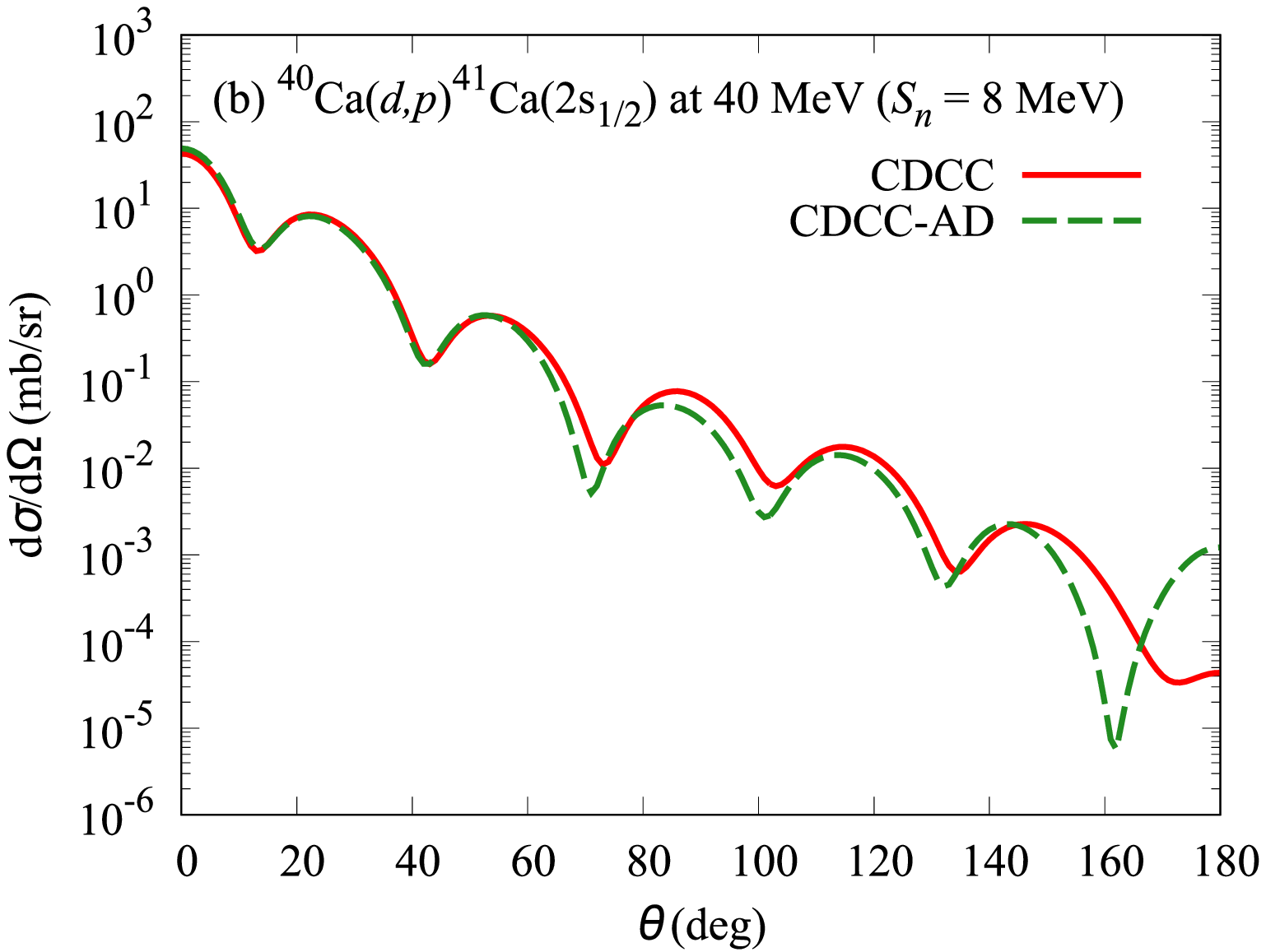}
 \caption{
(Color online) Angular distributions of the
$(d,p)$ cross sections calculated with
CDCC (solid lines) and CDCC-AD (dashed lines) for
(a) $^{100}$Zr$(d,p){}^{101}$Zr$(2s_{1/2})$ at $E_d=5$~MeV with $S_n=0.1$~MeV and
(b) $^{40}$Ca$(d,p){}^{41}$Ca$(2s_{1/2})$ at $E_d=40$~MeV with $S_n=8$~MeV.
  }
 \label{fig2}
\end{center}
\end{figure}
Figure~\ref{fig2}(a) shows the result for
$^{100}$Zr($d,p$)${}^{101}$Zr($2s_{1/2}$) at $E_d=5$~MeV
with $S_n=0.1$~MeV. The solid and dashed lines show
the results of CDCC and CDCC-AD, respectively.
When $E_d$ is much smaller than the Coulomb barrier height,
as is well known, the angular distribution is dictated by the
property of the Coulomb trajectory~\cite{Sat83} and has
a backward-peak structure; this is called Coulomb-dominated
transfer angular distributions.
In the case shown in Fig.~\ref{fig2}(a),
$S_{\rm AD}$ is 0.96 and the reaction can be regarded
as adiabatic. At the first look, it
seems to be strange that the AD approximation works at
such low incident energy. The reason for this is given
below in comparison with the result for the $S_n=8$~MeV case.

In Fig.~\ref{fig2}(b) we show the result for
$^{40}$Ca($d,p$)${}^{41}$Ca($2s_{1/2}$) at 40~MeV
and $S_n=8$~MeV. The incident energy is well above the
Coulomb barrier and the angular distribution shows the
diffraction pattern. The AD factor in this case is 0.87,
which shows the success of the AD approximation with about
10~\% error.  This is quite natural because as $E_d$
increases the deuteron internal motion becomes slow relative
to the motion of the c.m. of the deuteron, resulting in
the validness of the AD approximation. In general, this
is the case for $E_d \ge 20$~MeV with $S_n=8$~MeV. One
should keep it in mind, however, that there exists
not so large but finite difference coming from the
use of the AD approximation.

\begin{figure}[htpb]
\begin{center}
 \includegraphics[width=0.42\textwidth]{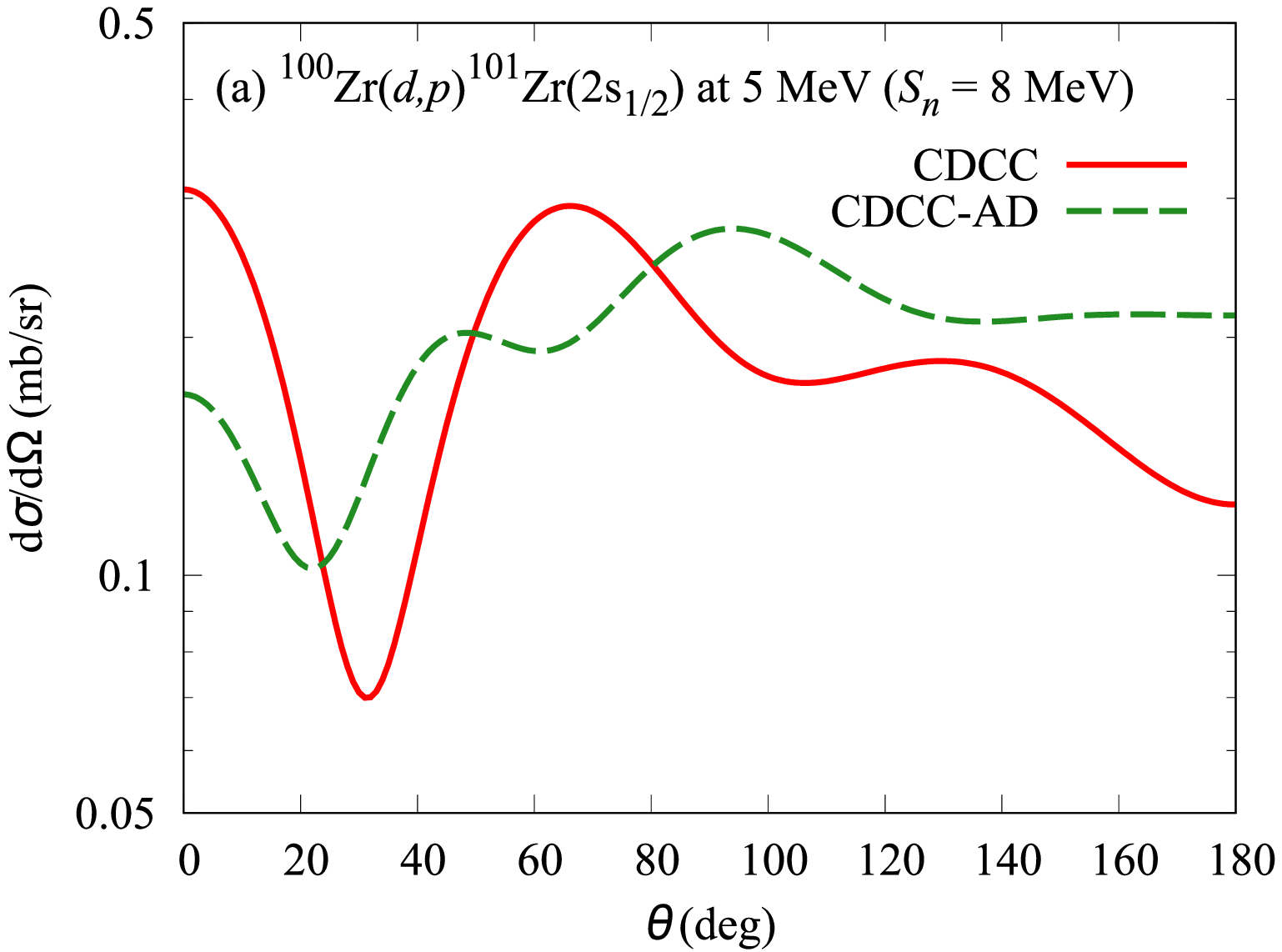}
 \includegraphics[width=0.42\textwidth]{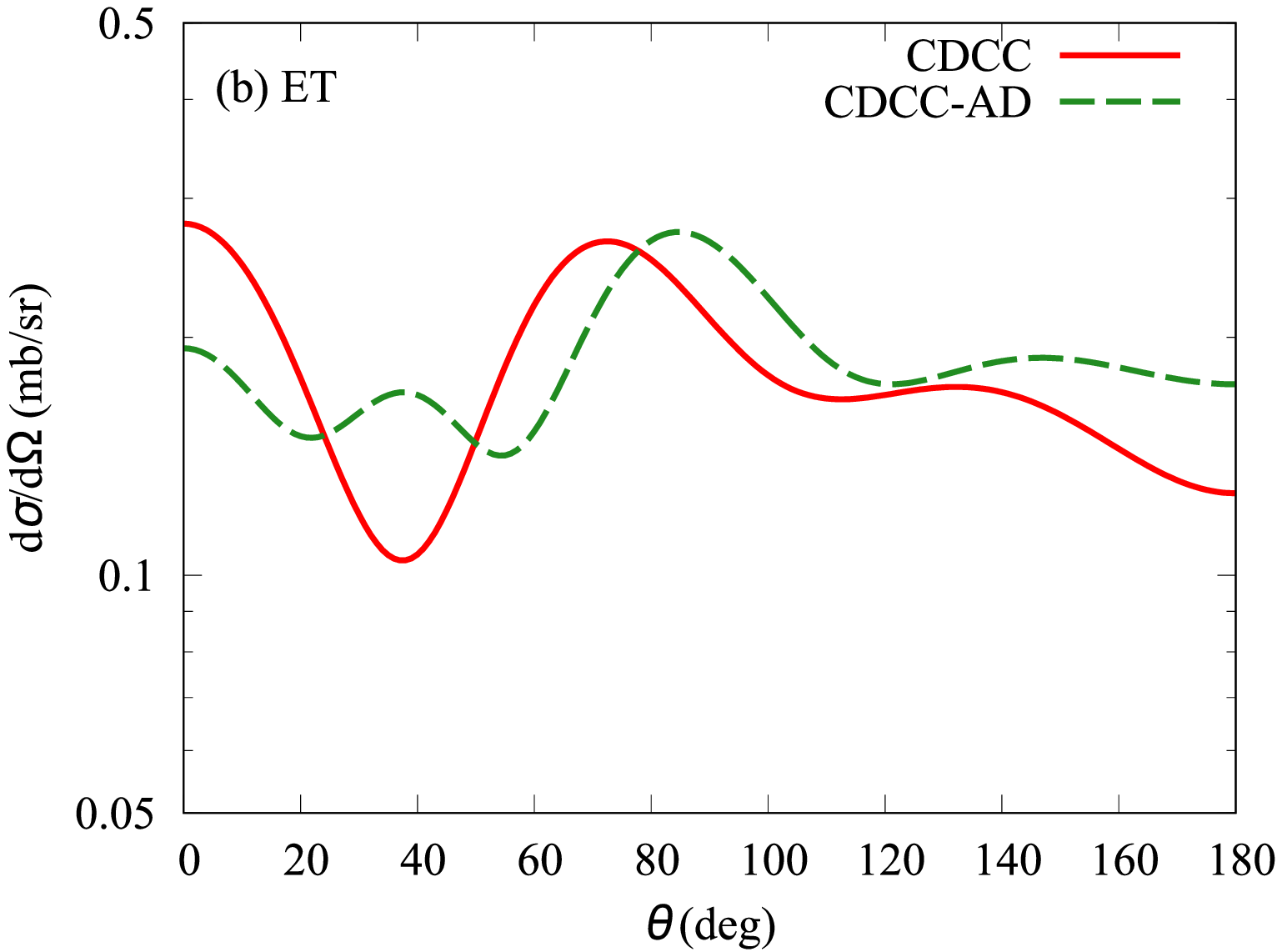}
 \includegraphics[width=0.42\textwidth]{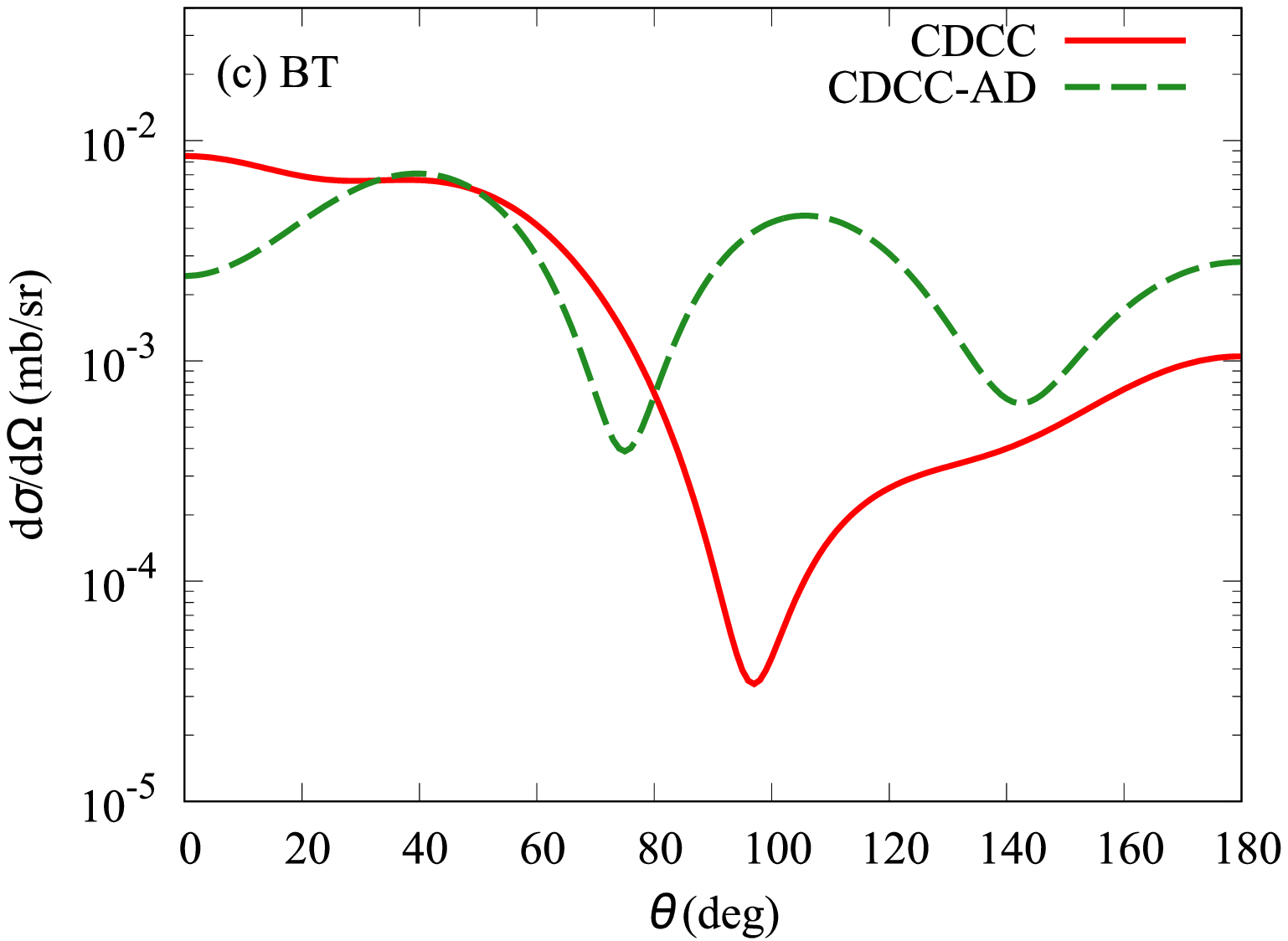}
 \caption{
(Color online) (a) Same as in Fig.~\ref{fig2} but for $S_n=8$ MeV;
(b) ET cross section and (c) BT cross section.
}
 \label{fig3}
\end{center}
\end{figure}
At lower energy, the validity of the AD approximation becomes
questionable. Although $S_{\rm AD}$ does not deviate from
unity very much, the ($d,p$) angular distribution seriously
suffers from the AD approximation for $E_d \le 10$~MeV and
$S_n=8$~MeV; we put *1 in Table~\ref{tab1} to specify the
systems for which this is the case.
As an typical example, the ($d,p$) cross section for
$^{100}$Zr($d,p$)${}^{101}$Zr($2s_{1/2}$) at $E_d=5$~MeV
with $S_n=8$~MeV is shown in Fig.~\ref{fig3}(a). Clearly,
the AD approximation fails to reproduce the result of
CDCC. In Figs.~\ref{fig3}(b) and~\ref{fig3}(c), we show
the cross sections of the ET and BT, respectively.
Despite the interference between the ET and BT amplitude
is not negligible, one may see that the difference between
the two lines in Fig.~\ref{fig3}(a) mainly comes from
that in the ET process. This suggests that the AD approximation
cannot treat the coupling of the breakup channels
to the elastic channel, that is, the so-called back-coupling.
The difference between the
two lines is very large also in the BT cross section.
Nevertheless, the BT process itself is not so important
because of its small contribution for the reaction systems
indicated by *1 in Table~\ref{tab1}.

\begin{figure}[htpb]
\begin{center}
 \includegraphics[width=0.42\textwidth]{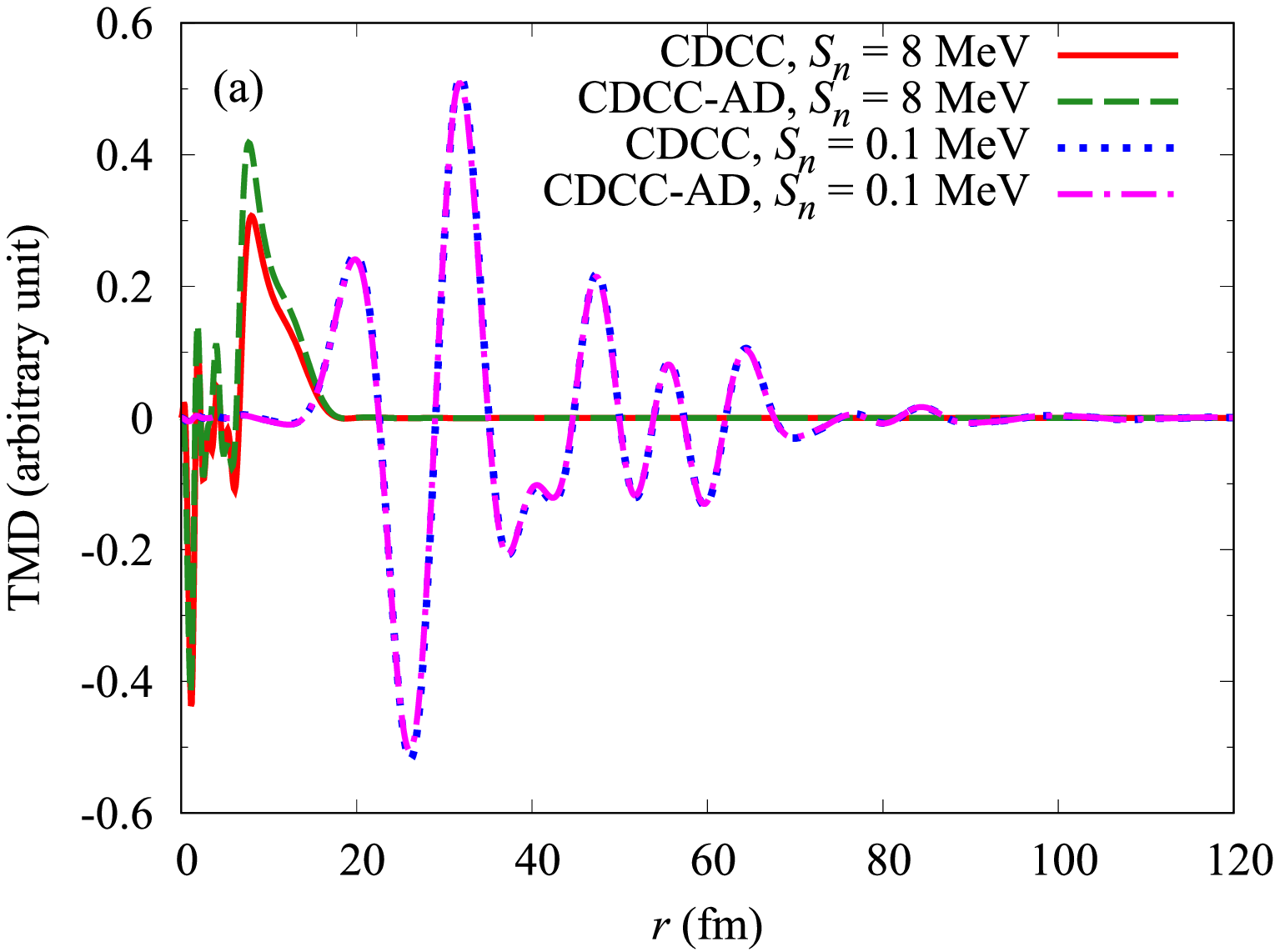}
 \includegraphics[width=0.42\textwidth]{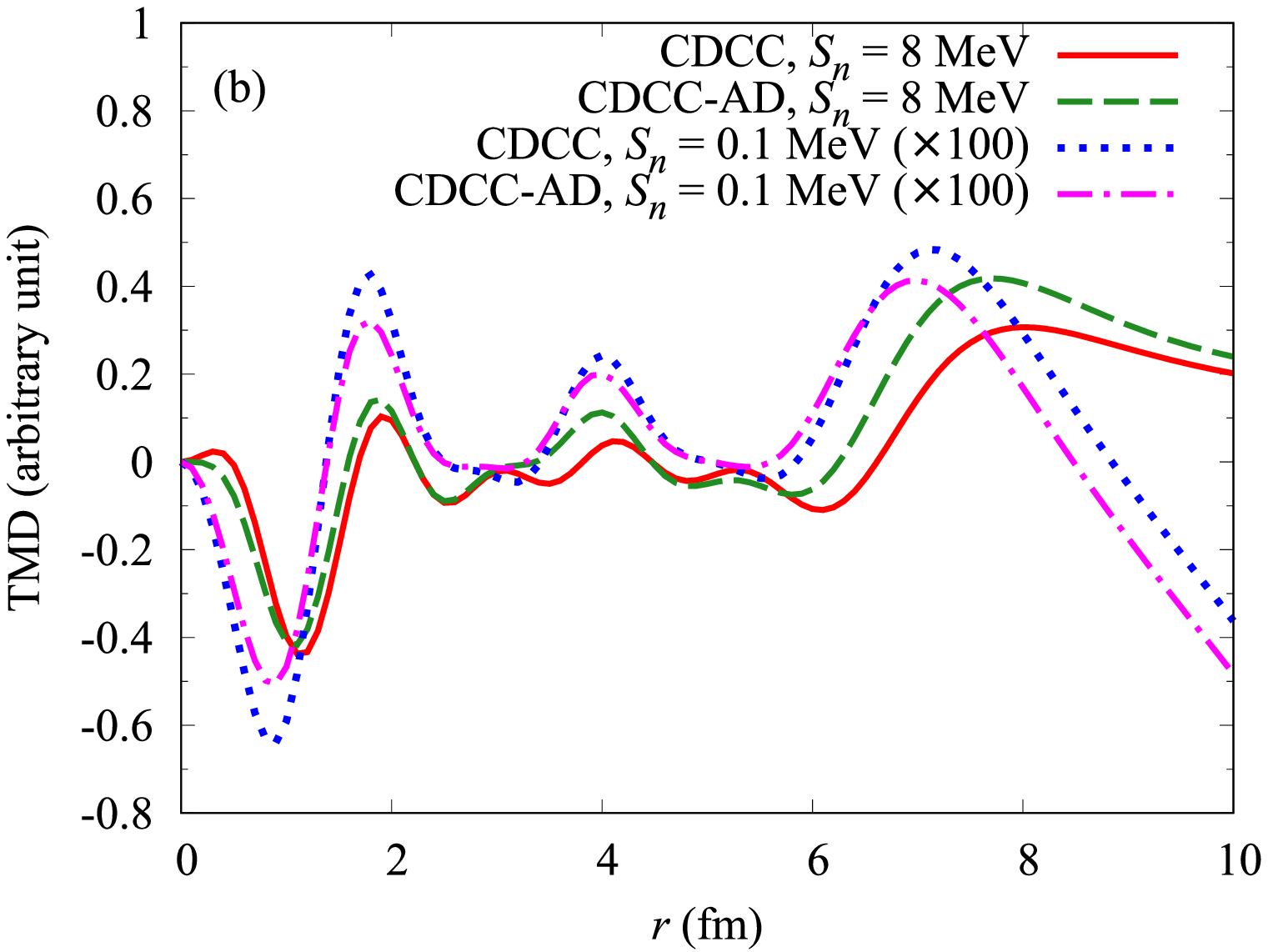}
 \caption{
(Color online) (a) TMDs for the ET part of the cross section
at $40^\circ$ for $^{100}$Zr$(d,p){}^{101}$Zr$(2s_{1/2})$ at $E_d=5$ MeV.
The solid (dashed) and dotted (dash-dotted) lines show the results with
CDCC (CDCC-AD) for $S_n=8$ and 0.1~MeV, respectively. (b) Enlarged view
of (a) for $r \le 10$~fm; the dotted and dash-dotted lines are multiplied
by 100.
  }
 \label{fig4}
\end{center}
\end{figure}
We discuss here the effect of $S_n$, which is the only
difference in the reaction systems shown in
Figs.~\ref{fig2}(a) and \ref{fig3}(a),
on the validity of the AD approximation. To see the role of
$S_n$ in more detail, we show in Fig.~\ref{fig4} the transition matrix
density (TMD) originally proposed in Ref.~\cite{Hat97} as a weighting
function for evaluating the mean density of the
($p,2p$) knockout reactions. The TMD can be interpreted as
a spatial distribution of the cross section; see
Refs.~\cite{Hat97,Nor98} for details.
The solid (dotted) and dashed (dash-dotted) lines in
Fig.~\ref{fig4}(a) show the TMDs for the ET cross section
at $\theta=40^\circ$
calculated with CDCC and CDCC-AD, respectively, for
$S_n=8$~MeV (0.1~MeV). One sees that the TMD for $S_n=0.1$~MeV
distributes from about 15~fm to 80~fm. In this region
the partial waves of $\Psi_\alpha^{(+)}$ for lower angular momenta
between A and the c.m. of the $p$-$n$ system, which
are distorted by $U_p$ and $U_n$, have only a small contribution
to $\Psi_\alpha^{(+)}$. In other words, the incident-wave part of $\Psi_\alpha^{(+)}$
is dominant there. The use of the AD approximation therefore
makes no difference in the ET amplitude. In fact, the breakup
effect itself is found to be negligibly small, which trivially results in
the tiny contribution of the BT process.
This is why CDCC-AD successfully reproduces the result of CDCC
for the reaction shown in Fig.~\ref{fig2}(a).
On the other hand,
the TMD distributes below about 15~fm when $S_n=8$~MeV.
In that region,
the nuclear distortion including the back-coupling
effect is significant. As mentioned, because of the low incident
energy, the breakup effect cannot be treated accurately by the
AD approximation.

As mentioned above, there is no
difference in $\Psi_\alpha^{(+)}$ for
Fig.~\ref{fig2}(a) and Fig.~\ref{fig3}(a).
What classifies the validity of the AD approximation is therefore
the region where the reaction takes place. If the nuclear
interior and surface regions are important, the AD approximation
fails at low incident energies. If only the tail (asymptotic)
region is important, the AD approximation works well even at low incident energies.
In Fig.~\ref{fig4}(b) the results for $r \le 10$~fm are shown;
those for $S_n=0.1$~MeV are multiplied by 100. One may see
the difference coming
from the AD approximation indeed exists also for $S_n=0.1$~MeV.
As mentioned, however, this region does not have a meaningful
contribution to the cross section, resulting in the
success of CDCC-AD.

\begin{figure}[htpb]
\begin{center}
 \includegraphics[width=0.42\textwidth]{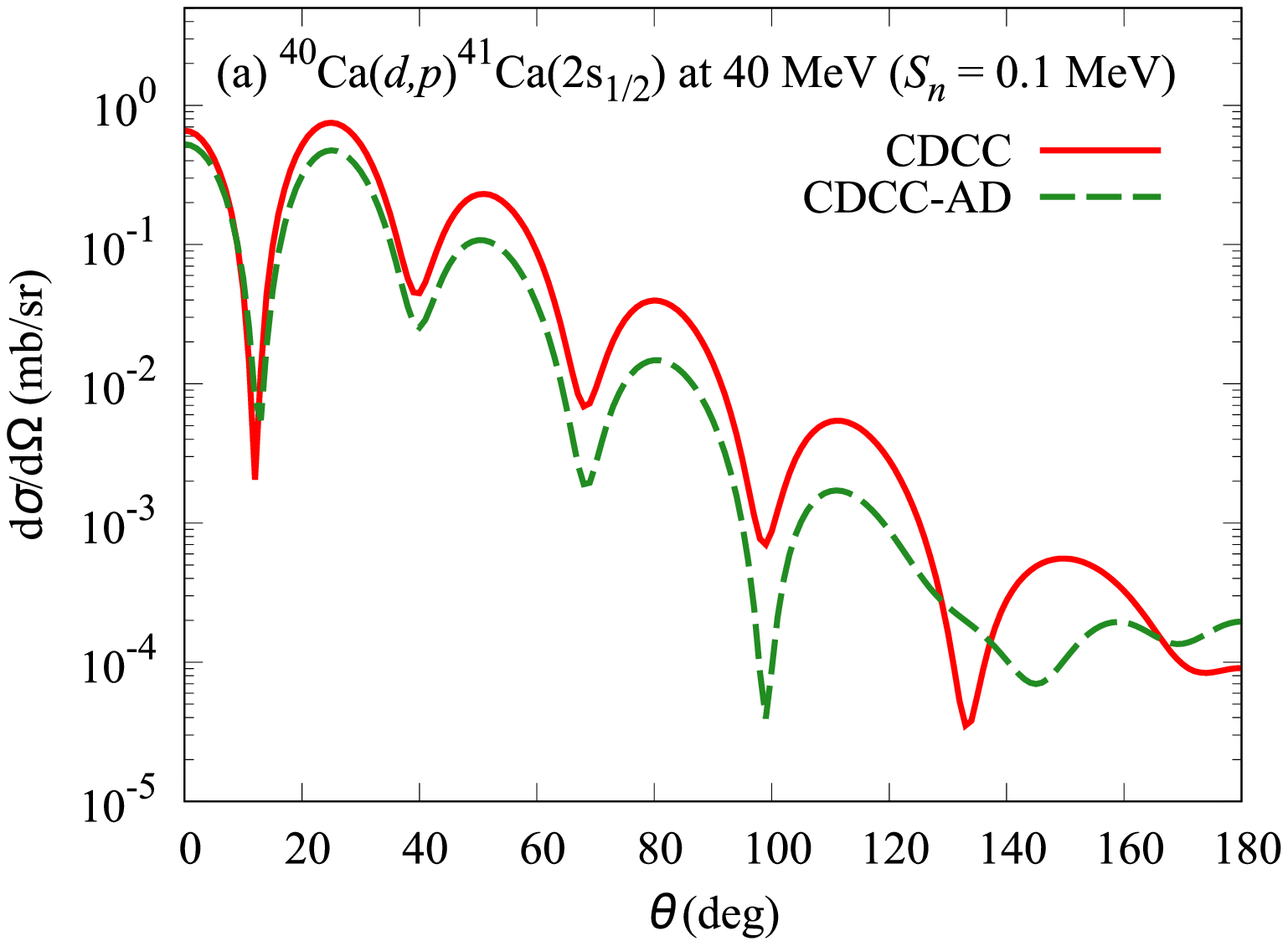}
 \includegraphics[width=0.42\textwidth]{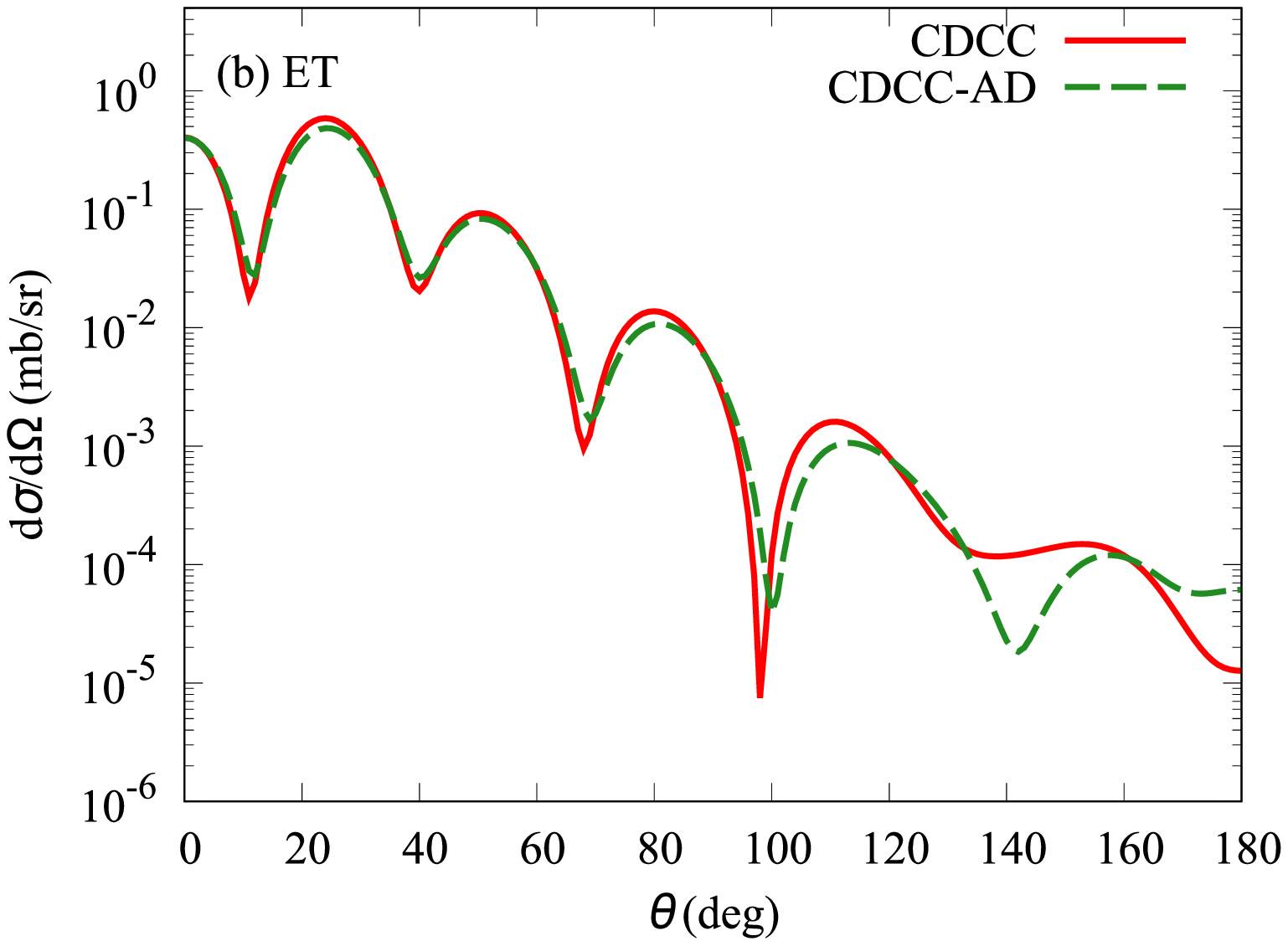}
 \includegraphics[width=0.42\textwidth]{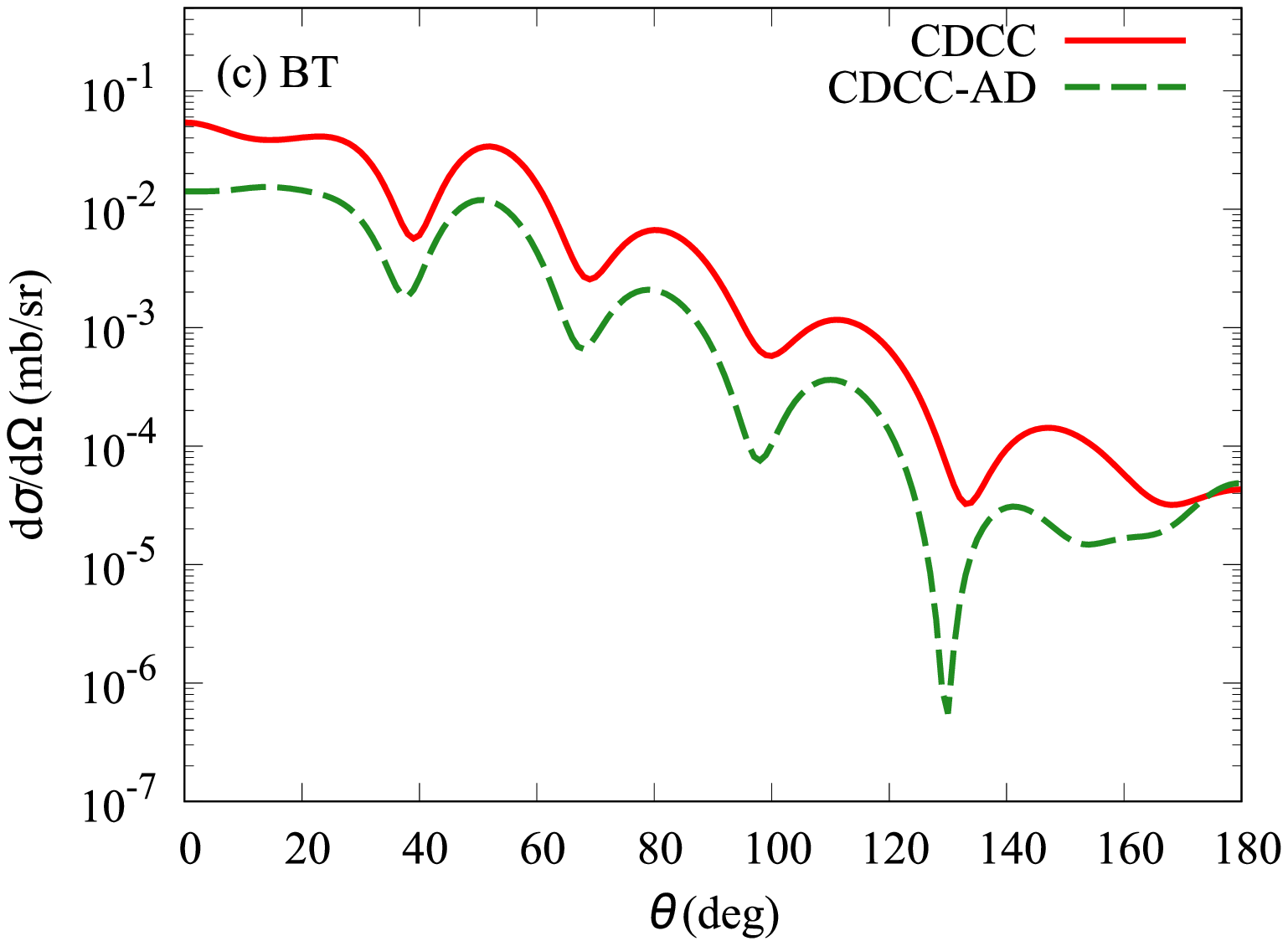}
 \caption{
(Color online) Same as in Fig.~\ref{fig3} but for
$^{40}$Ca$(d,p){}^{41}$Ca$(2s_{1/2})$ at $E_d=40$ MeV with $S_n=0.1$ MeV.
  }
 \label{fig5}
\end{center}
\end{figure}
Next we discuss the cases for which $S_{\rm AD}$ is
significantly large; we put *2 in Table~\ref{tab1}
for them. Figure~\ref{fig5}(a) shows the result for
$^{40}$Ca($d,p$)${}^{41}$Ca($2s_{1/2}$) at 40~MeV
with $S_n=0.1$~MeV, and Figs.~\ref{fig5}(b) and \ref{fig5}(c)
show the corresponding ET and BT cross sections, respectively.
In this case, the result of CDCC-AD undershoots that of CDCC
for the BT part, whereas
the two calculations give almost the same result
for the ET cross section except at very backward angles.
Thus, in some cases for $S_n=0.1$~MeV and at relatively
high incident energies, the AD approximation fails to describe
the breakup property of deuteron in the ($d,p$) process.
In consequence of this, the absolute value of the
cross section calculated with CDCC-AD significantly
smaller than that of CDCC.

\begin{figure}[htpb]
\begin{center}
 \includegraphics[width=0.42\textwidth]{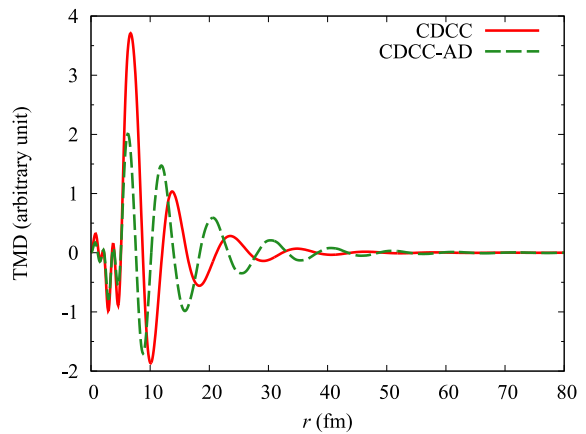}
 \caption{
(Color online) TMD for the BT cross section in Fig.~\ref{fig5}(c) at $0^\circ$.
  }
 \label{fig6}
\end{center}
\end{figure}
It is well known that the AD approximation tends to overshoot
the breakup cross section of the projectile, since
the AD approximation makes all the $p$-$n$ continuum
states degenerate to the g.s. of deuteron,
and thus makes the coupling between the deuteron g.s. and
its breakup states effectively stronger. In fact, the deuteron elastic breakup
cross section $\sigma_{\rm EB}$
calculated with CDCC-AD is 107~mb
and that with CDCC is 73~mb. On the other hand,
the result of CDCC-AD is smaller than that of CDCC for
the BT cross section as mentioned above.
To see this in more detail, we show in Fig.~\ref{fig6}
the TMD for the BT cross section corresponding to $\theta=0^\circ$.
In the tail region, the amplitude of the CDCC-AD is larger than
that of CDCC, reflecting mainly the amplitudes of the deuteron
scattering wave function in the breakup channels.
This is consistent with the aforementioned results of
$\sigma_{\rm EB}$. On the other hand,
in the surface region around 7~fm, the result of CDCC (the solid line)
has a larger positive value than that of CDCC-AD (the dashed line).
Since the integrated value of
the TMD is proportional to the cross section, the
larger BT cross section of CDCC shown in
Fig.~\ref{fig5}(c)
is due to the behavior of the solid line
in Fig.~\ref{fig6} around 7~fm. It is,
however, difficult
to pin down the reason for this internal behavior,
mainly because of the complicated coupled-channel effects.

\begin{figure}[htpb]
\begin{center}
 \includegraphics[width=0.42\textwidth]{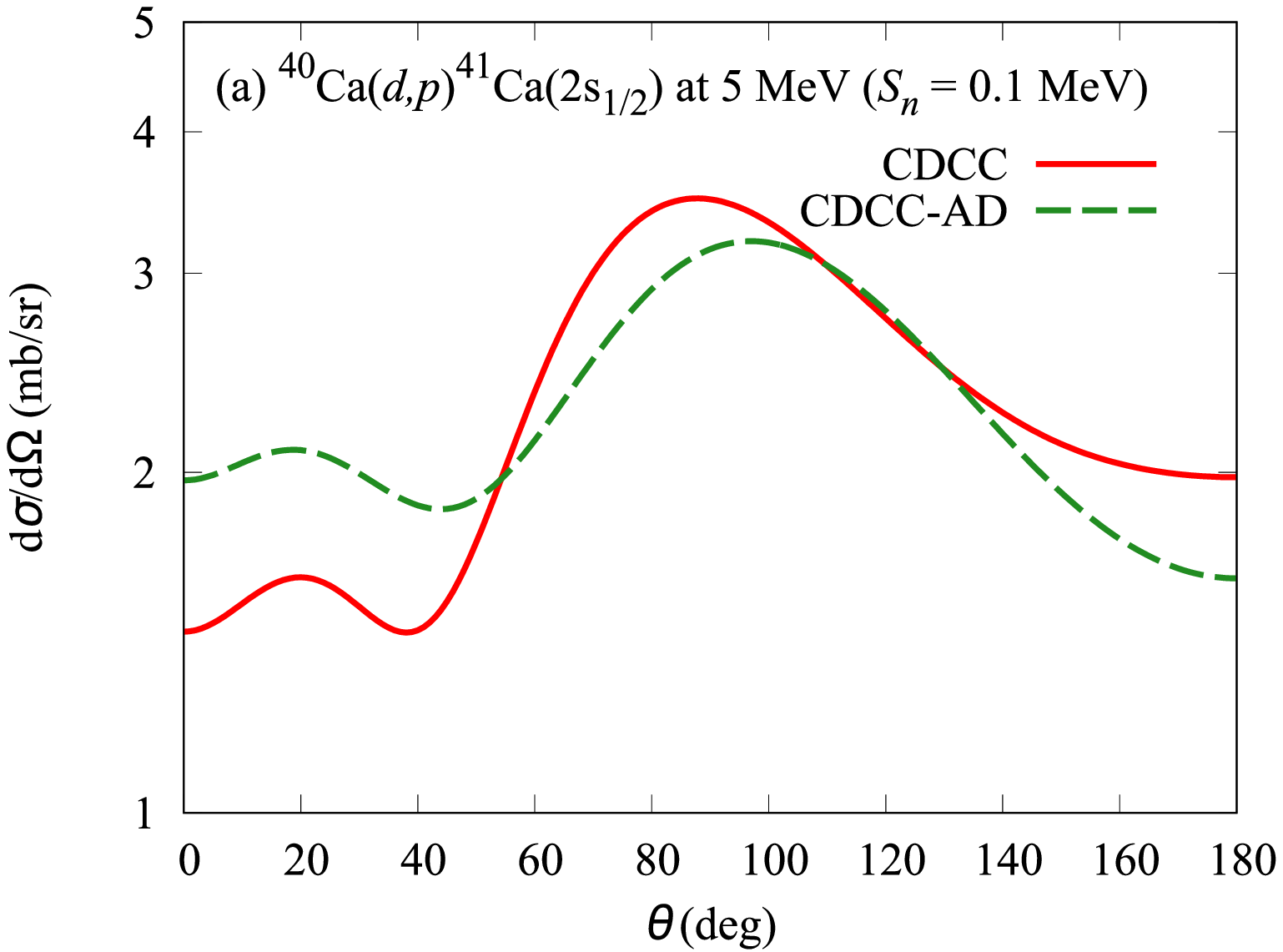}
 \includegraphics[width=0.42\textwidth]{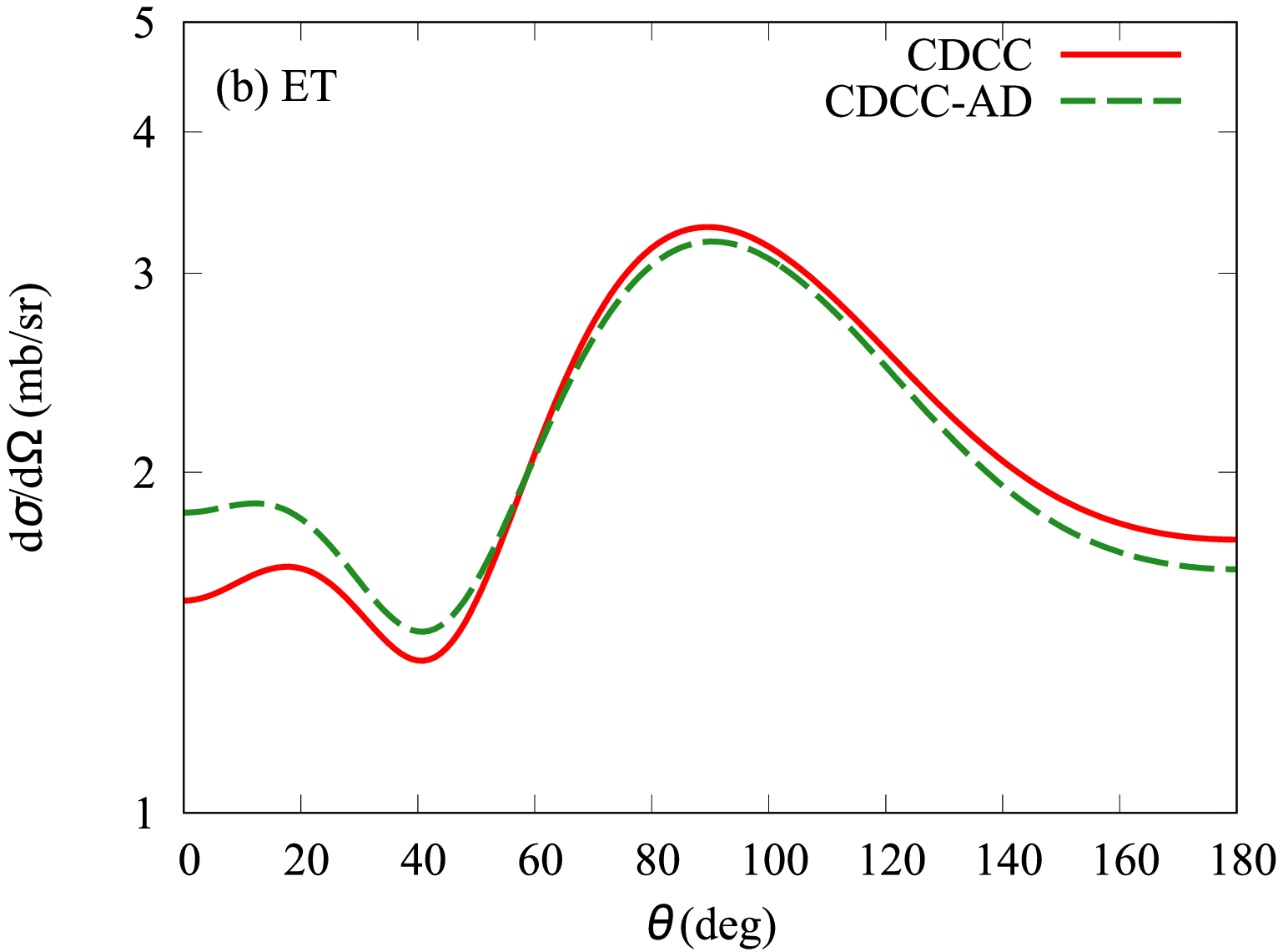}
 \includegraphics[width=0.42\textwidth]{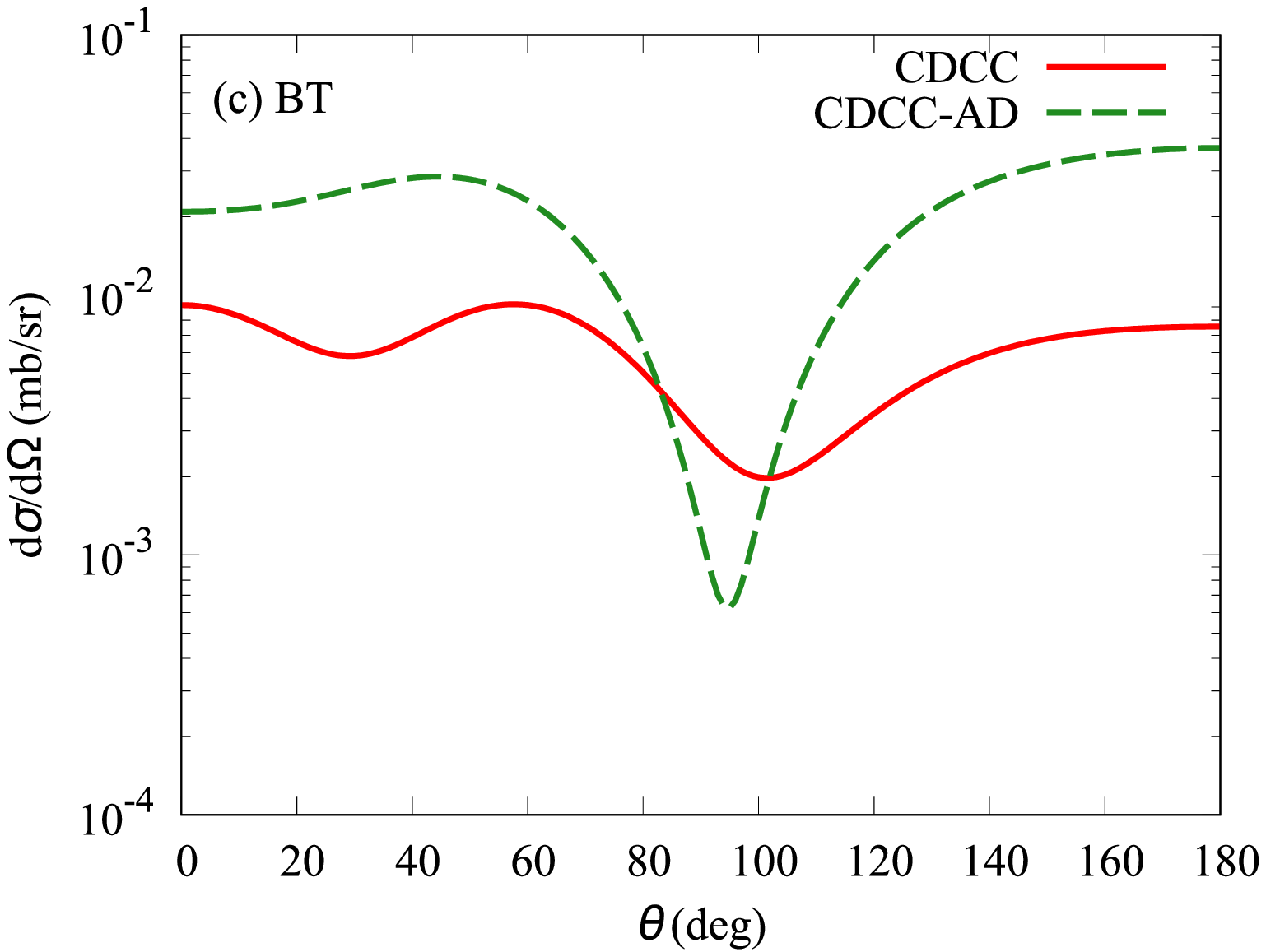}
 \caption{
(Color online) Same as in Fig.~\ref{fig3} but for
$^{40}$Ca$(d,p){}^{41}$Ca$(2s_{1/2})$ at $E_d=5$ MeV with $S_n=0.1$ MeV.
  }
 \label{fig7}
\end{center}
\end{figure}
Finally, we discuss the cases
in which $S_n=0.1$~MeV and the result of CDCC-AD
deviates from that of CDCC, even though $S_{\rm AD}\sim 1$;
we put *3 for them in Table~\ref{tab1}. The $(d,p)$ cross sections
of $^{40}$Ca($d,p$)${}^{41}$Ca($2s_{1/2}$) at 5~MeV with $S_n=0.1$~MeV
are shown in Fig.~\ref{fig7}, as in Fig.~\ref{fig3}.
The shape of the cross section of CDCC-AD is somewhat
different from that of CDCC, which is attributed to the
difference in the BT cross sections. One of the important
characteristics of this reaction system is the relation
between $E_d$ and the Coulomb barrier height $V_{\rm CB}$.
When $E_d \ll V_{\rm CB}$, the Coulomb-dominated
transfer angular distributions are observed, whereas
the diffraction pattern develops when $E_d > V_{\rm CB}$~\cite{Sat83}.
There is a window for $E_d$ between these two
conditions, that is, $E_d \sim V_{\rm CB}$. In this region,
the shape of the cross section starts changing from the
Coulomb-dominated distribution to the diffraction pattern.
The balance between $E_d$ and $V_{\rm CB}$ is thus crucially
important there. In CDCC, when the incident deuteron breaks up,
the energy of the c.m. motion of the $p$-$n$ system decreases
following the energy conservation of the three-body system.
When $E_d \sim V_{\rm CB}$, the $p$-$n$ c.m. energy
in the breakup channels goes below $V_{\rm CB}$, and the BT
hardly contributes to the $(d,p)$ cross section because of
the Coulomb barrier. On the contrary,
the AD approximation ignores the energy conservation and the
penetrability of the scattering wave in breakup channels
is the same as in the incident channel. As a result,
the BT cross section is significantly overestimated by the
CDCC-AD calculation. This is the case when $E_d \sim V_{\rm CB}$
and $S_n=0.1$~MeV; in fact, a similar result is obtained for
$^{200}$Hg($d,p$)${}^{201}$Hg($3s_{1/2}$) around 15~MeV with $S_n=0.1$~MeV.
When $S_n$ is large, say, 8~MeV, the contribution of the BT
becomes less important and the validity of the AD approximation
mainly relies on the accurate description of the ET process
as mentioned above.

Thus far we have discussed the validity of the AD approximation
with respect to $E_d$, $S_n$, and target nuclei. As for the
trend in $\ell_f$, one can conclude from Table~\ref{tab1} that
when $S_n=8$~MeV the selectivity of $\ell_f$ is weak and
$E_d$ dictates the accuracy of the AD approximation.
On the other hand,
for $S_n=0.1$~MeV
almost all the nonadiabatic cases are
found when $\ell_f=0$; this may be
related to the halo structure of the $n$-A system.

\subsection{Effect of the closed channel}
\label{sec33}

\begin{table}[hbtp]
 \caption{Values of $S_{\rm OP}$. The superscript $\dagger$ represents the cases in which
$S_{\rm OP}$ does not deviate much from unity but
the angular distribution is severely affected by the neglect of the closed channels.
}
 \label{tab2}
 \centering
 \begin{tabular}{ccccccccccc}

  \multicolumn{11}{c}{ } \\
  \multicolumn{11}{l}{$ell_f=0$} \\
  \hline \hline
  & & \multicolumn{4}{c}{Energy ($S_n=0.1$ MeV)} & & \multicolumn{4}{c}{Energy ($S_n=8$ MeV)} \\
  Target & & 5 & 10 & 20 & 40 & & 5 & 10 & 20 & 40 \\
  \hline
  $^{20}$Ne & & 1.00 & 1.09 & 1.16 & 1.00 & & 0.85 & 1.04 & 1.10 & 0.91 \\
  $^{40}$Ca & & 1.27 & 1.39 & 1.25 & 0.95 & & 0.58 & 0.92 & 1.04 & 0.92 \\
  $^{100}$Zr & & 1.00 & 1.18$^{\dagger}$ & 1.13 & 0.94 & & 0.89$^{\dagger}$ & 1.02 & 0.84 & 0.93 \\
  $^{200}$Hg & & 1.00 & 0.99 & 0.94$^{\dagger}$ & 0.94 & & 1.08$^{\dagger}$ & 0.84$^{\dagger}$ & 0.89 & 0.96 \\
  \hline \hline

  \multicolumn{11}{c}{ } \\
  \multicolumn{11}{l}{$ell_f=1$} \\
  \hline \hline
  & & \multicolumn{4}{c}{Energy ($S_n=0.1$ MeV)} & & \multicolumn{4}{c}{Energy ($S_n=8$ MeV)} \\
  Target & & 5 & 10 & 20 & 40 & & 5 & 10 & 20 & 40 \\
  \hline
  $^{20}$Ne & & 1.13 & 1.07 & 1.02 & 0.97 & & 0.90 & 1.04 & 1.00 & 0.96 \\
  $^{40}$Ca & & 1.19$^{\dagger}$ & 0.99 & 0.95 & 0.99 & & 0.62 & 0.88 & 0.97 & 0.97 \\
  $^{100}$Zr & & 1.00 & 1.02$^{\dagger}$ & 0.91 & 0.95 & & 0.69 & 0.88 & 0.89 & 0.98 \\
  $^{200}$Hg & & 1.00 & 0.99 & 0.98 & 0.95 & & 0.94 & 0.72 & 0.89 & 0.98 \\
  \hline \hline

  \multicolumn{11}{c}{ } \\
  \multicolumn{11}{l}{$ell_f=2$} \\
  \hline \hline
  & & \multicolumn{4}{c}{Energy ($S_n=0.1$ MeV)} & & \multicolumn{4}{c}{Energy ($S_n=8$ MeV)} \\
  Target & & 5 & 10 & 20 & 40 & & 5 & 10 & 20 & 40 \\
  \hline
  $^{20}$Ne & & 1.04 & 1.02 & 0.95 & 1.00 & & 0.79 & 0.91 & 0.93 & 1.00 \\
  $^{40}$Ca & & 1.06$^{\dagger}$ & 0.96 & 0.96 & 0.97 & & 0.79 & 0.91 & 0.94 & 0.98 \\
  $^{100}$Zr & & 1.01 & 0.96 & 0.98 & 0.95 & & 0.92$^{\dagger}$ & 0.94 & 0.90 & 1.00 \\
  $^{200}$Hg & & 1.00 & 0.99 & 0.95 & 0.96 & & 1.05 & 0.89$^{\dagger}$ & 0.90 & 1.00 \\
  \hline \hline

  \multicolumn{11}{c}{ } \\
  \multicolumn{11}{l}{$ell_f=3$} \\
  \hline \hline
  & & \multicolumn{4}{c}{Energy ($S_n=0.1$ MeV)} & & \multicolumn{4}{c}{Energy ($S_n=8$ MeV)} \\
  Target & & 5 & 10 & 20 & 40 & & 5 & 10 & 20 & 40 \\
  \hline
  $^{20}$Ne & & 0.93$^{\dagger}$ & 0.96 & 0.98 & 0.99 & & 1.13$^{\dagger}$ & 1.01 & 1.02 & 1.00 \\
  $^{40}$Ca & & 1.07$^{\dagger}$ & 0.92 & 0.96 & 0.98 & & 0.69 & 0.77 & 0.92 & 1.00 \\
  $^{100}$Zr & & 1.01 & 0.97 & 0.94 & 0.98 & & 0.80$^{\dagger}$ & 0.71 & 0.89 & 0.99 \\
  $^{200}$Hg & & 1.00 & 0.99 & 0.96 & 1.00 & & 1.00 & 0.98$^{\dagger}$ & 0.90 & 1.00 \\
  \hline \hline

 \end{tabular}
\end{table}

In our CDCC calculation, as mentioned, the maximum $p$-$n$ linear momentum
$k_{\rm max}$ is taken to be 2.0~fm$^{-1}$.
In some studies, however, $k_{\rm max}$ is determined by
\beq
\frac{\hbar^2 k^2_{\rm max}}{2\mu_{pn}}=E_0,
\label{kop}
\eeq
where $\mu_{pn}$ is the reduced mass of the $p$-$n$ system
and $E_0$ is the deuteron incident energy in the c.m. system.
In other words, the so-called closed channels are sometimes neglected.
Recently, it was found that the inclusion of the closed channels
in CDCC is crucial for accurately describing the deuteron
breakup cross sections at low incident energies~\cite{OY16}.

\begin{figure}[htpb]
\begin{center}
 \includegraphics[width=0.42\textwidth]{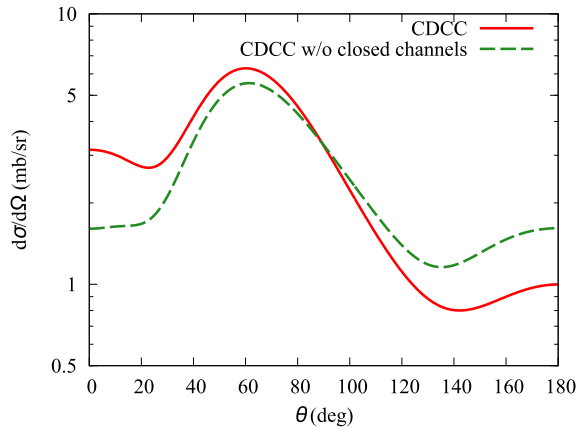}
 \caption{
(Color online) Angular distributions of the
$(d,p)$ cross sections for
$^{20}$Ne$(d,p){}^{21}$Ne$(0f_{7/2})$ at $E_d=5$~MeV with $S_n=8$~MeV.
The solid and dashed lines show the results of CDCC with and without
the closed channels, respectively.
  }
 \label{fig8}
\end{center}
\end{figure}

To see the importance of the closed channels for the $(d,p)$ processes,
we show in Table~\ref{tab2} the factor $S_{\rm OP}$ defined
in the same way as for $S_{\rm AD}$ but with
$d\sigma_{\rm AD}/d\Omega$ in Eq.~(\ref{sad}) replaced with
$d\sigma_{\rm OP}/d\Omega$; $d\sigma_{\rm OP}/d\Omega$
is the result of CDCC with $k_{\rm max}$ determined by Eq.~(\ref{kop}).
As expected, for $E_d \ge 20$~MeV the closed channels have no
significant effect, resulting in $S_{\rm OP}\sim 1$.
However, in some cases the neglect of
the closed channels affects the result by more than 10~\% even in
that energy region.
At lower energy, the effect of the closed channels can be
very large, for $S_n=8$~MeV in particular.
Furthermore, for the reaction systems indicated by
$\dagger$ in Table~\ref{tab2}, neglect of the closed channels
significantly changes the angular distribution,
even though $S_{\rm OP}$ does not different from unity.
Figure~\ref{fig8} shows a typical example for those cases.

\begin{figure}[htpb]
\begin{center}
 \includegraphics[width=0.42\textwidth]{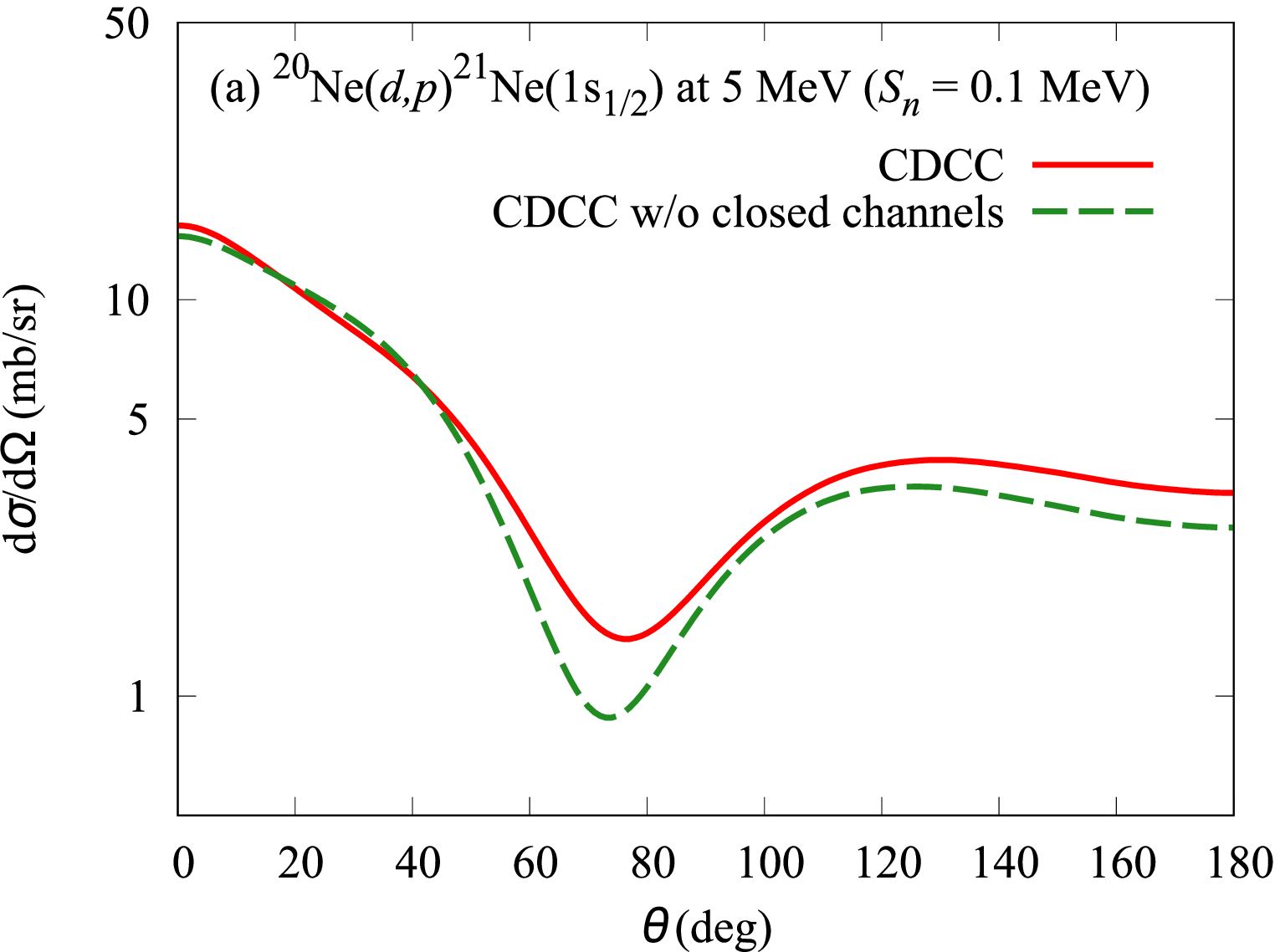}
 \includegraphics[width=0.42\textwidth]{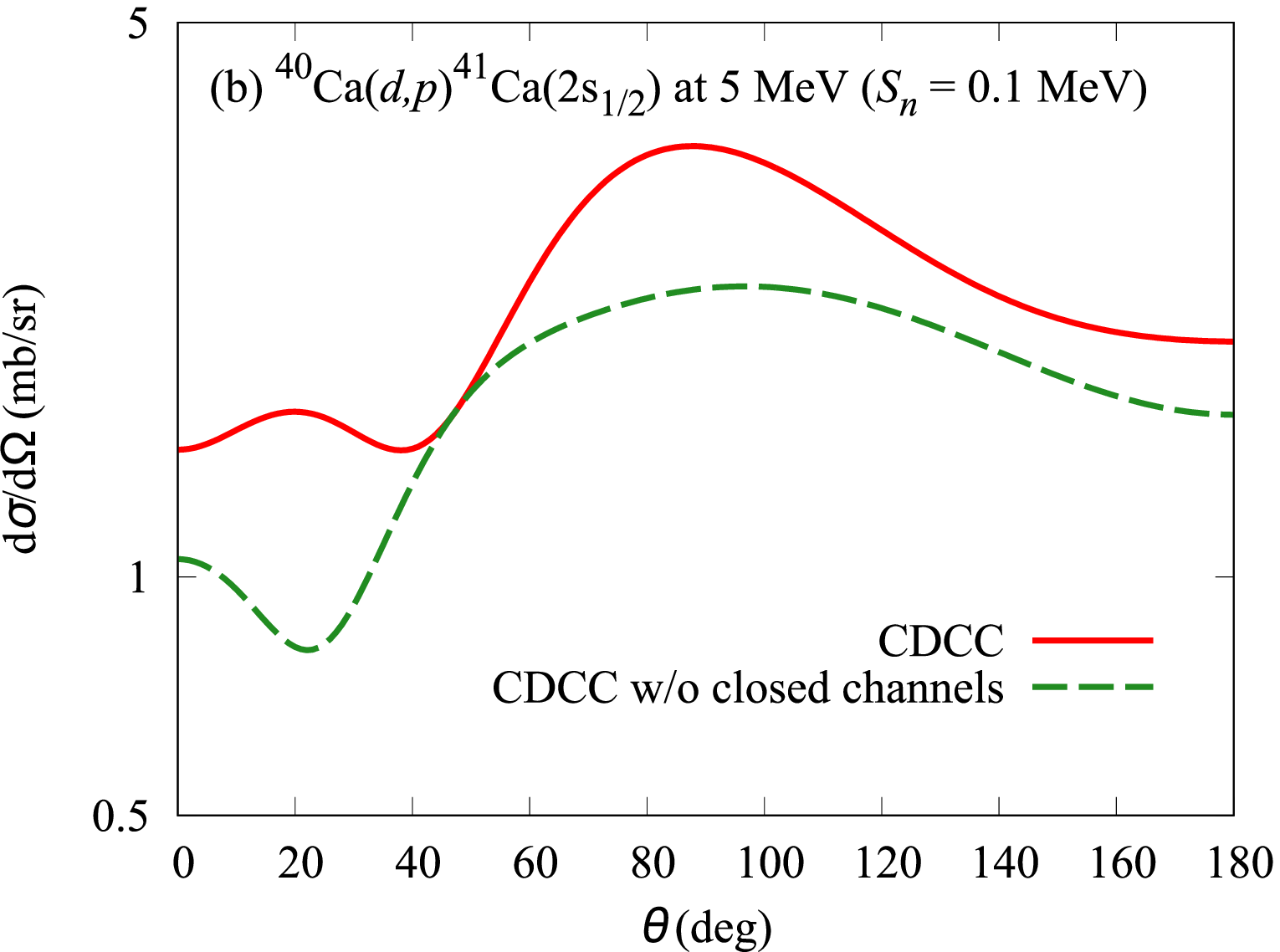}
 \includegraphics[width=0.42\textwidth]{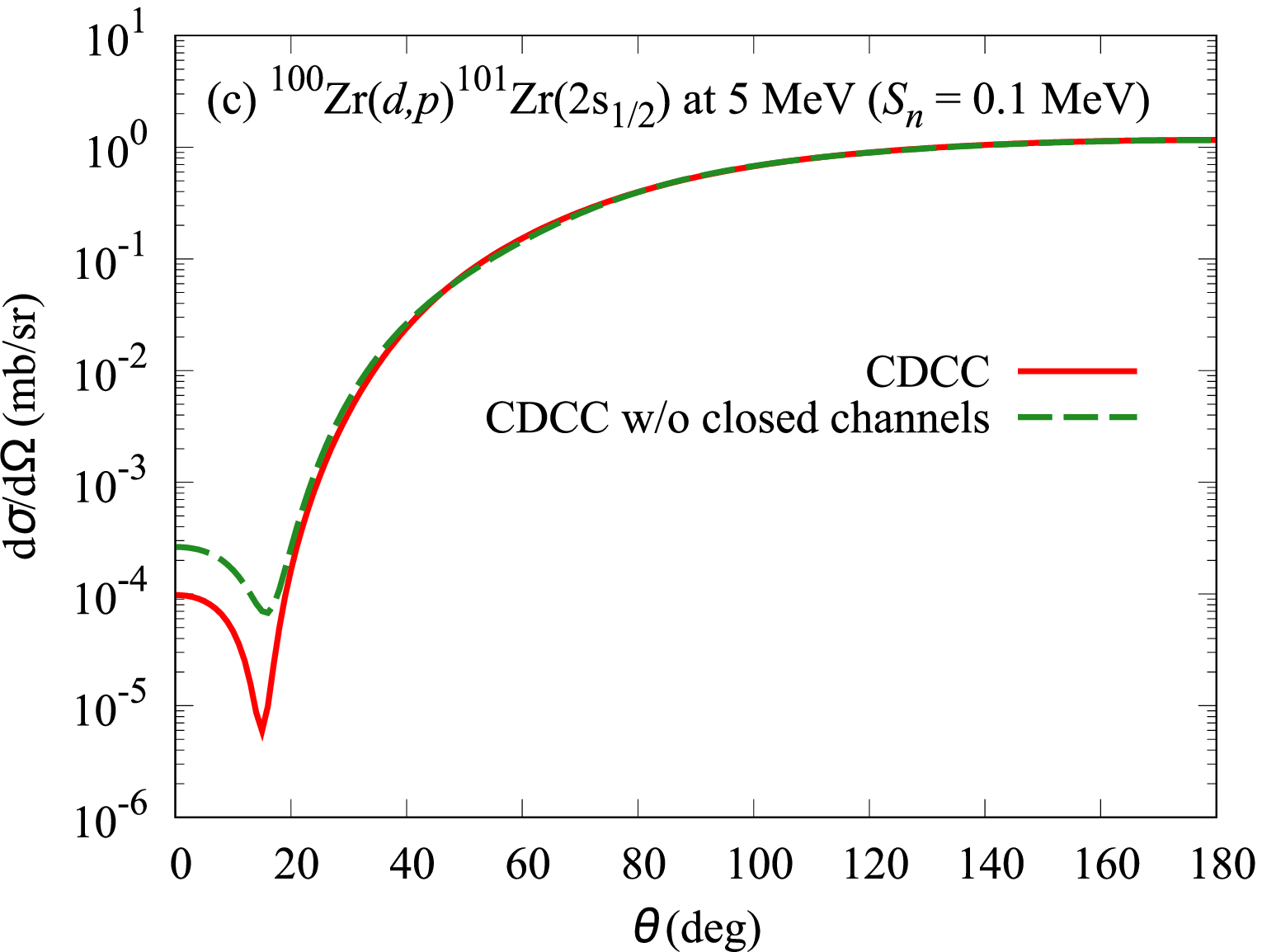}
 \caption{
(Color online) Same as in Fig.~\ref{fig8} but for
(a) $^{20}$Ne$(d,p){}^{21}$Ne$(0f_{7/2})$,
(b) $^{40}$Ca$(d,p){}^{41}$Ca$(2s_{1/2})$,
and
(c) $^{100}$Zr$(d,p){}^{101}$Zr$(2s_{1/2})$,
at $E_d=5$ MeV with $S_n=0.1$ MeV.
  }
 \label{fig9}
\end{center}
\end{figure}

By taking a closer look at Table~\ref{tab2},
one may find that the tendency of $S_{\rm OP}$
is quite nontrivial. For instance, when $\ell_f=0$, $S_n=0.1$~MeV,
and $E_d=5$~MeV, $S_{\rm OP}$ significantly deviates from unity
only for $^{40}$Ca. In Fig.~\ref{fig9} we show the results of
comparison for $^{20}$Ne, $^{40}$Ca, and $^{100}$Zr.

%

One sees that a strikingly large effect of the closed channels
appears when $E_d \sim V_{\rm CB}$, as in the *3 cases mentioned
in Sec.~\ref{sec32}. For $\ell_f \neq 0$, however, this seems
not to be the case. Thus, we conclude that
it is difficult to see {\it a priori} the role of the closed channels.
We thus conclude that the use of Eq.~(\ref{kop})
is not recommended; $k_{\rm max}$ must be determined so as
to make the physics observables calculated with CDCC converged.
Comparison for all the reaction systems as in Fig.~\ref{fig8}
can be found in the addendum provided as supplemental material~\cite{SM17}.

\section{Summary}
\label{sec4}

We have examined the validity of the adiabatic (AD) approximation
to the deuteron-target three-body wave function in the
calculation of the cross section of the ($d,p$) process
for 128 reaction systems. For this purpose, results of CDCC
that explicitly treat the breakup channels are compared
with those of CDCC with the AD approximation (CDCC-AD).
The typical error due to the AD approximation
is found to be less than 20~\% and around 35~\% at most.
However, there are three exceptional cases in which
the AD approximation does not work.

First, when the deuteron incident energy $E_d$ is
less than 10~MeV and the neutron separation energy
$S_n$ in the residual nucleus is 8~MeV, the AD approximation
cannot describe the ($d,p$) angular distribution calculated
by CDCC, mainly because of the failure in describing
the elastic transfer process. This will be natural because
the assumption of the AD approximation, that is,
the assumption that the internal motion of deuteron is much slower than
that of the c.m. of deuteron, does not hold.
In this case, however, if an appropriate optical potential
that can describe the deuteron elastic channel is provided,
the ($d,p$) process does not suffer from the deuteron breakup
effect.

Second, for some reaction systems in which $E_d \ge 20$~MeV
and $S_n=0.1$~MeV, the result of CDCC-AD is significantly smaller
than that of CDCC. We found that this is due to the undershooting
of the breakup transfer contribution by the AD approximation.
It should be noted that the AD approximation overshoots
the deuteron breakup cross section, because it enhances
the breakup probability of deuteron in general. The effect
due to the AD approximation on the breakup transfer (BT) process
is opposite to it and will be a consequence of complicated
coupled-channel effects.

Third, when $E_d$ is close to the Coulomb barrier energy
and $S_n=0.1$~MeV, the behavior of the BT process cannot
be properly described by the AD approximation, because
it violates the energy conservation of the three-body system;
the energy of the c.m. of the $p$-$n$ system does not
change even after breakup and can penetrate the Coulomb
barrier as in the elastic channel.

We have investigated also the effect of the closed channels.
For $E_d \le 20$~MeV, the neglect of the closed channels
can seriously affect the result, for $S_n=8$~MeV in
particular. However, there seems no clear threshold above
which the  closed channels can be neglected. It will be
recommended that the convergence of the CDCC model
space with respect to $k_{\rm max}$ should always be
confirmed, as for other quantities such as $l_{\rm max}$
and $\Delta_k$.

In this study the energy dependence and nonlocality of the
distorting potential as well as the finite-range effect in
the $(d,p)$ process are not discussed. Moreover, the breakup
effect in the final channel is not taken into account.
The findings summarized
above therefore will need further investigation in view of these
additional aspects. A more complete analysis will be very important.

\section*{Acknowledgments}

The authors thank K.~Minomo and Y.~S.~Neoh for fruitful discussions.
The numerical calculation was carried out with the computer
facilities at the Research Center for Nuclear Physics, Osaka
University. This work was supported in part by Grants-in-Aid
of the Japan Society for the Promotion of Science (Grants No.
JP15J01392 and No. JP25400255)
and by the ImPACT Program of the Council for Science, Technology
and Innovation (Cabinet Office, Government of Japan).


\end{document}